\documentclass[5p,twocolumn,10pt,number]{elsarticle}
\usepackage{pifont}
\usepackage{natbib}
\usepackage{geometry}
\usepackage{fleqn}
\usepackage{hyperref}
\usepackage{epsfig}
\usepackage{graphicx}
\usepackage{bm}
\usepackage{amssymb,amsmath}
\usepackage{revsymb}
\usepackage{graphics}
\usepackage[usenames,dvipsnames]{color}
\usepackage{nomencl}
\makenomenclature
\journal{Journal of Quantitative Spectroscopy and Radiative Transfer}
\begin{document}

\begin{frontmatter}

\title{A Green's function formalism of energy and momentum transfer in fluctuational electrodynamics}

\author[Columbia]{A. Narayanaswamy\corref{cor1}}
\ead{arvind.narayanaswamy@columbia.edu}

\author[Columbia]{Y. Zheng}

\address[Columbia]{Department of Mechanical Engineering, Columbia University, New York, NY 10027}

\cortext[cor1]{Corresponding author}

\date{\today}

\begin{abstract}
Radiative energy and momentum transfer due to fluctuations of electromagnetic fields arising due to temperature difference between objects is described in terms of the cross-spectral densities of the electromagnetic fields. We derive relations between thermal non-equilibrium contributions to energy and momentum transfer and surface integrals of tangential components of the dyadic Green's functions of the vector Helmholtz equation. The expressions derived here are applicable to objects of arbitrary shapes, dielectric functions, as well as magnetic permeabilities. For the case of radiative transfer, we derive expressions for the generalized transmissivity and generalized conductance that are shown to obey reciprocity and agree with theory of black body radiative transfer in the appropriate limit.  
\end{abstract}


\end{frontmatter}             

\printnomenclature
\begin{thenomenclature}
\item[{$\mathbf{E}$}] {Electric field vector}
\item[{$F$}] {View factor }
\item[{$\overline{\overline{G}}_e$}] {Electric dyadic Green's function }
\item[{$\overline{\overline{G}}_m$}] {Magnetic dyadic Green's function }
\item[{$\overline{\overline{G}}_E$}] {$\nabla  \times \overline{\overline{G}}_e$ }
\item[{$\overline{\overline{G}}_M$}] {$\nabla \times \overline{\overline{G}}_m$ }
\item[{$\overline{\overline{G}}_o$}] {Green's function of contribution due to background or source radiation }
\item[{$\overline{\overline{G}}^{(sc)}$}] {Green's function of contribution from waves scattered by  interfaces }
\item[{$G^e$}] {Linearized conductance for radiative transfer }
\item[{$\mathbf{H}$}] {Magnetic field vector}
\item[{$\overline{\overline{I}}$}] {Identity matrix }
\item[{$\mathbf{J}$}] {Current Density}
\item[{$T_l$}] {Temperature in object $l$ }
\item[{$\mathbf{P}$}] {Poynting vector }
\item[{$Q$}] {Radiative heat transfer }
\item[{$\tilde{R}_{hn}$}] {Fresnel reflection coefficients at interfaces between $h$ and $n$ }
\item[{$S_l$}] {Closed surface of object $l$ }
\item[{$T^e$}] {Generalized transmissivity for radiative energy transfer }
\item[{$T^m$}] {Generalized transmissivity for momentum transfer }
\item[{$V_l$}] {Volume of object $l$ }
\item[{$V_\delta$}] {Volume of infinitesimal radius surrounding $\mathbf{\tilde r}$ }
\item[{$\overline{\overline{\mathcal{E}}}$}] {Matrix of contribution to $\langle \mathbf{E}    \mathbf{E}^*  \rangle_s$ }
\item[{$\overline{\overline{\mathcal{H}}}$}] {Matrix of contribution to $\langle \mathbf{H}    \mathbf{H}^*  \rangle_s$ }
\item[{$\overline{\overline{\mathcal{X}}}$}] {Matrix of contribution to $\langle \mathbf{E}    \mathbf{H}^*  \rangle_s$ }

\item[{$c$}] {Speed of light }
\item[{$\hbar$}] {Reduced Planck's constant }
\item[{$k$}] {Wavevector }
\item[{$k_b$}] {Boltzmann's constant }
\item[{$k_n$}] {n component of wavevector ($n=x,y,z$) }
\item[{$k_{hz}$}] {$z$ component of wavevector in vacuum }
\item[{$k_\rho$}] {$\sqrt{k_x^2+k_y^2}$ }
\item[{$l$}] {Thickness of vacuum gap }
\item[{$\mathbf{\hat n}$}] {Unit normal vector }
\item[{$\mathbf{r}$}] {Position vector of observation point}
\item[{$\mathbf{\tilde{r}}$}] {Position vector of source point}
\item[{$t$}] {Time}
\item[{$\beta$}] {$0$ or $1$ }
\item[{$\delta$}] {Delta function }
\item[{$\mathbf{\rho}$}] {Distance of points}
\item[{$\epsilon$}] {Levi-Civita symbol }
\item[{$\varepsilon$}] {Permittivity, $\varepsilon'+i\varepsilon''$ }
\item[{$\varepsilon_o$}] {Permittivity of free space }
\item[{$\Theta$}] {Energy of a photon at temperature T }
\item[{$\mu$}] {Permeability, $\mu'+i\mu''$ }
\item[{$\mu_o$}] {Permeability of free space }
\item[{$\nu,\xi$}] {$1$ or $-1$ }
\item[{$\overline{\overline{\sigma}}$}] {Maxwell stress tensor }
\item[{$\omega$}] {Frequency}
\item[{$\Re$}] {Real part }
\item[{$\Im$}] {Imaginary part }
\item[{$Tr$}] {Trace}

\item[\textbf{Superscripts}]{}
\item[{$bb$}] {Blackbody }
\item[{$e$}] {Electric field }
\item[{$m$}] {Magnetic field }
\item[{$pp$}] {Planar-planar }
\item[{$(h)$}] {Vacuum }
\item[{$(l)$}] {Objects ($l=1,2,\cdots,N$) }
\item[{$(p)$}] {Transverse magnectic }
\item[{$(s)$}] {Transverse electric }
\item[{$(\mu)$}] {Polarization $s$ or $p$ }
\item[{$T$}] {Transpose }
\item[{$*$}] {Complex conjugate }

\item[\textbf{Subscripts}]{}
\item[{$h$}] {Vacuum }
\item[{$i,p,q$}] {Cartesian components $1,2,3$ }
\item[{$l$}] {Objects ($l=1,2,\cdots,N$) }
\item[{$s$}] {Symmetric summation}
\item[{$1\rightarrow 2$}] {From object 1 to 2 }
\end{thenomenclature}

\section{Introduction}

Fluctuations of electromagnetic fields lead to thermal radiative transfer, via energy transfer, and van der Waals  and Casimir forces, via momentum transfer. Diffraction and interference effects as well as tunneling of evanescent and surface waves, collectively known as near-field effects, are not taken into consideration by the classical theory of radiative transfer. Near-field effects become important when the length scale of importance becomes comparable to the characteristic thermal wavelength ($ \lambda_T \approx 3000/T $ $ \mu $m). For radiative transfer between two objects, an important length scale is the minimum inter-object spacing, $ l_{gap} $. When $ l_{gap} \ll \lambda_T $, tunneling of electromagnetic waves lead to enhancement of radiative transfer beyond the classical or far-field limit.  Surface texturing, for instance by creating a periodic 1D or 2D pattern, introduces a length scale, $ l_{p} $, that characterizes the period of the pattern.  When $ l_{p} \ll \lambda_T $, diffraction effects can lead to thermal emission patterns not usually associated with a planar surface \cite{greffet02a}. 

 It has been long recognized that near-field enhancement of radiative transfer due to surface polaritons can result in increased power density as well as efficiency \cite{narayanaswamy03a,laroche06b,basu07,basu2009review}. However, this enhancement of energy transfer has not been used in any practical device, as yet, because of our inability to conceive of configurations other than two parallel surfaces with a thin vacuum gap in which an enhancement of similar magnitude occurs. Most investigations of near-field radiative transfer have been restricted to objects of few simple geometric shapes, each analyzed by a vector eigenfunction expansion method applicable to that geometry (planar geometry with vector plane waves \cite{narayanaswamy03a,francoeur2008near,francoeur2008role,francoeur2009solution,biehs2010mesoscopic,ben2010fundamental}, cylindrical surfaces with vector cylindrical waves \cite{kruger2011nonequilibrium}, two spheres with vector spherical waves \cite{narayanaswamy07b,sasihithlu2011convergence,sasihithlu2011proximity,carrillo2010nanosphere}, sphere-plane with a combination of vector spherical and plane waves \cite{otey2011numerically}). Even minor changes to the shape of the object can impose great challenges. Simulations of thermal emission from textured surfaces are usually performed using rigorous coupled wave analysis (RCWA) \cite{wang2011phonon,wang2012wavelength,zhang2011measurements}or finite difference time domain (FDTD)  methods \cite{rodriguez2011frequency}, which are quite different from those used for simulations of near-field radiative transfer.  To design other types of surfaces that can exploit the enhancement, without posing the hurdles associated with two parallel surfaces, and  also to design surfaces with new radiative properties by shape modification at nano/micro scale, we need a general method to predict all types of nanoscale effects on radiative transfer, irrespective of the size, shape or properties of the objects involved. 

Kruger et al. \cite{kruger2011nonequilibrium,kruger2012trace} used fluctuational electrodynamics to develop a scattering matrix and operator formalism for computing non-equilibrium force and heat transfer interactions between objects with arbitrary shapes and frequency dependent dielectric permittivities. Biehs et. al. \cite{biehs2010mesoscopic} developed a formalism of nanoscale radiative transfer between two parallel surfaces similar to that of Landauer formalism of electron transport in mesoscopic devices \cite{landauer1957spatial,landauer1989conductance,imry1999conductance,datta1997electronic}.  Ben-Abdallah et. al. \cite{ben2011many} used Rytov's theory to develop a theoretical formalism for radiative transfer between many objects in the dipole limit. Messina et al. \cite{messina2011scattering} proposed a scattering matrix version of nanoscale radiative transfer as well as dispersion forces that is valid for objects with arbitrary shapes as well as dielectric functions. Non-equilibrium fluctuational electrodynamical interactions between objects can be expressed in a scattering matrix formalism or in a Green's functions formalism, just as the electrical conductance for electron transport can be developed in terms of the scattering matrix or Green's function. 

The work in this paper is an extension to a prior work published in this journal by one of the authors \cite{narayanaswamy2010a}. In Ref. \cite{narayanaswamy2010a}, the focus was on the relation between cross-spectral densities of electromagnetic fields in thermal equilibrium and the dyadic Green's functions (DGFs) of the vector Helmholtz equation. In this paper, the focus is on thermal non-equilibrium effects, i.e. when the objects are at different temperatures. Volume integral expressions for cross-spectral densities of components of the electric and magnetic fields that can be obtained from Rytov's theory of fluctuational electrodynamics are converted into a form more appropriate (in terms of surface integrals of DGFs of the vector Helmholtz equation) for computations as well as comparison with the classical theory of radiative transfer. Though the focus of this paper is not on developing new numerical techniques, it is hoped that the formalism developed here will be used to compute thermal non-equilibrium energy and momentum transfer between arbitrarily shaped objects. 



The paper is arranged as follows. In Sec. \ref{sec:fegff}, the fluctuation-dissipation theorem and DGFs are used to express the electric and magnetic field correlation functions in terms of the volume integrals of the DGFs. In Sec. \ref{sec:fesif}, Green's identities for dyadic functions are used to derive expressions for the field correlations in terms of surface integrals of tangential components of the DGFs. 
 For radiative transfer between two objects, a generalized transmissivity function that is expressed in terms of double surface integrals on the surfaces of the two objects is derived in Sec. \ref{sec:generalizedtransmissivity}. In Sec. \ref{sec:specgeom}, the theoretical formalism developed in Sec. \ref{sec:generalizedtransmissivity} is applied to the cases of  radiative heat transfer between two parallel half spaces. We also show that the generalized transmissivity function agrees with the theory of blackbody radiation in the appropriate limit. Finally, we also discuss the implications of the theoretical formalism developed here for computation of heat transfer and non-equilibrium forces. 

\begin{figure}
\center
\includegraphics[width=8.5cm]{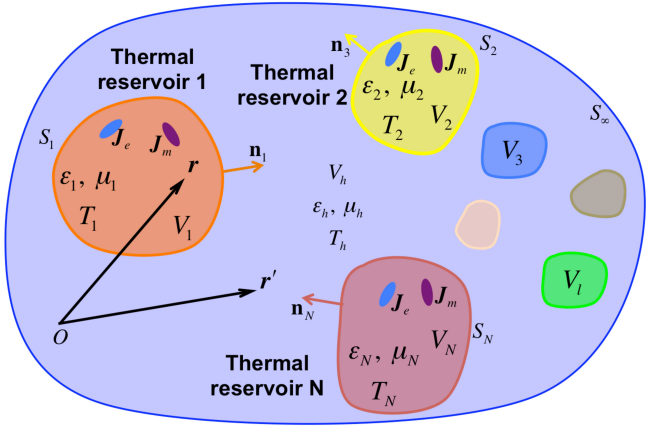}
\caption{\label{fig:generalinteraction}Schematic of $ N $ objects at temperatures $T_1, T_2,\cdots,T_N$ embedded in a host medium at $T_{h}$. } 
\end{figure}

\section{\label{sec:fegff}Fluctuational electrodynamics and Green's function formalism}

We briefly describe our notation regarding electromagnetic fields and their Fourier transforms here. 
A field $ A(\mathbf{r},t) $ and its Fourier transform, $ A(\mathbf{r},\omega) $, are related by $ A(\mathbf{r},t)=\frac{1}{2\pi}\int\limits_{-\infty}^{\infty} A(\mathbf{r},\omega)e^{-i\omega t} d\omega $. Since the same symbol is used to identify a field as well as its Fourier transform, explicit dependence on time will be included when refering to the time domain field. 
Explicit dependence on $ \omega $ is suppressed from $ A(\mathbf{r},\omega) $ so that it is written as $ A(\mathbf{r}) $. Explicit dependence of relative dielectric permittivities, magnetic permeabilities, and DGFs on $ \omega $ is also suppressed. 

Let us consider $ N $ objects (see Fig. \ref{fig:generalinteraction}) with relative dielectric permittivities $\varepsilon_l(\omega)$ and magnetic permeabilities $\mu_l(\omega)$ at temperatures $T_l$, where $ l=1,2,\cdots,N $. These objects are assumed to be embedded in vacuum that is at temperature $T_{h}$. The object $ l $ is confined to the volume $ V_l $ and the closed surface $ S_l $ is the boundary of this object with the host medium. The outward normal on the surface of object $ l $ at $ \mathbf{r} $ is represented by $ \mathbf{n}_l (\mathbf{r}) $. 
The fluctuations of the electric and magnetic current densities, which give rise to the dispersion forces and radiative transfer, are related to temperature by fluctuation-dissipation theorems of the \textit{second kind} \cite{callen51a,landau69a,eckhardt1982first}:
\begin{subequations}
\label{eqn:fdresults}
\begin{eqnarray}
\label{eqn:jejestrength}
\langle J^{e}_{p}(\mathbf{r})J^{e*}_{q}(\mathbf{\tilde{r}}) \rangle &=&2\omega\varepsilon_{o}\varepsilon''\Theta\left(\omega,T\right)\delta(\mathbf{r}-\mathbf{\tilde{r}})\delta_{pq} \\
\label{eqn:jmjmstrength}
\langle J^{m}_{p}(\mathbf{r})J^{m*}_{q}(\mathbf{\tilde{r}}) \rangle &=& 2\omega\mu_{o}\mu''\Theta\left(\omega,T\right)\delta(\mathbf{r}-\mathbf{\tilde{r}})\delta_{pq}\\
\label{eqn:jejmstrength}
\langle J^{e}_{p}(\mathbf{r})J^{m*}_{q}(\mathbf{\tilde{r}}) \rangle &=& 0
\end{eqnarray}
\end{subequations}

\noindent where $p,q=1,2,3$ are the labels for the Cartesian components of the vector, $ \varepsilon_o $ and $ \mu_o $ are the permittivity and permeability of free space, $J^{e}_p$ and $J^{m}_{p}$ are the Cartesian components of the electric and magnetic current densities, $\Theta\left(\omega,T\right) =\displaystyle \frac{\hbar\omega}{2}\coth\left(\frac{\hbar\omega}{2k_b T}\right)$, $\varepsilon''$ and $\mu''$ are the imaginary parts of the dielectric permittivity and magnetic permeability respectively at the location $\mathbf{r}$ which is in local thermodynamic equilibrium at temperature $T$,  $ z^* $ is the complex conjugate of $ z $, and $ \langle \rangle $ denotes the ensemble average. $ 2\pi\hbar  $ is the Planck constant and $ k_b $ is the Boltzmann constant. The presence of $\delta_{pq}$ implies that we assume all materials to be isotropic, and that of $\delta(\mathbf{r}-\mathbf{\tilde{r}})$ implies that the correlations of sources are local. The Fourier transforms of the electric and magnetic fields in the host medium (volume $V_{h}$ in Fig. \ref{fig:generalinteraction}) due to sources in object $ l $ (volume $V_1$) are given by:
\begin{subequations}
\label{eqn:fieldrelations}
\begin{equation}
\label{eqn:efieldeqn}
 \mathbf{E}(\mathbf{\tilde{r}})  =\int\limits_{V_l} [ \mathbf{p} (\mathbf{r})  \cdot \overline{\overline{\mathbf{G}}}_{e}(\mathbf{r},\mathbf{\tilde{r}})  - \mathbf{J}^{m}(\mathbf{r}) \cdot \overline{\overline{\mathbf{G}}}_{E}(\mathbf{r}, \mathbf{\tilde{r}}) ]d\mathbf{r}
 \end{equation} 
 \begin{equation}
\label{eqn:mfieldeqn}
\mathbf{H}(\mathbf{\tilde{r}})= \int\limits_{V_l} [\mathbf{m} (\mathbf{r}) \cdot \overline{\overline{\mathbf{G}}}_{m}(\mathbf{r}, \mathbf{\tilde{r}}) +  \mathbf{J}^{e}(\mathbf{r}) \cdot \overline{\overline{\mathbf{G}}}_{M}(\mathbf{r}, \mathbf{\tilde{r}}) ] d\mathbf{r}
\end{equation}
\end{subequations}

\noindent where $\mathbf{p} \left(\mathbf{r}\right)$$=$$i\omega\mu_{o}\mu(\mathbf{r})\mathbf{J^{e}}\left(\mathbf{r}\right)$, $\mathbf{m} \left(\mathbf{r}\right) = i\omega\varepsilon_{o}\varepsilon(\mathbf{r}) \mathbf{J^{m}}\left(\mathbf{r}\right)$, $\mathbf{\tilde{r}} \in V_{h}$, $\overline{\overline{\mathbf{G}}}_{E}(\mathbf{r},\mathbf{\tilde{r}})=\nabla \times  \overline{\overline{\mathbf{G}}}_{e}(\mathbf{r},\mathbf{\tilde{r}})$ and $\overline{\overline{\mathbf{G}}}_{M}(\mathbf{r},\mathbf{\tilde{r}}) =\nabla \times  \overline{\overline{\mathbf{G}}}_{m}(\mathbf{r},\mathbf{\tilde{r}})$. $\overline{\overline{\mathbf{G}}}_{e}(\mathbf{r},\mathbf{\tilde{r}}) $ and $\overline{\overline{\mathbf{G}}}_{m}(\mathbf{r},\mathbf{\tilde{r}}) $ are DGFs of the vector Helmholtz equation that satisfy the following boundary conditions on the interface $ S_l $ between object $ l $ and the host medium:
\begin{subequations}
\begin{equation}
\label{eqn:ebc}
\hat{\mathbf{n}}_l(\mathbf{r}_l) \times ( \mu_l(\mathbf{r}_l) \overline{\overline{\mathbf{G}}}_{e}(\mathbf{r}_l,\mathbf{\tilde{r}})-\mu_{h}(\mathbf{r}_h) \overline{\overline{\mathbf{G}}}_{e}(\mathbf{r}_{h},\mathbf{\tilde{r}})) = 0, 
\end{equation}
\begin{equation}
\label{eqn:Ebc}
\hat{\mathbf{n}}_l(\mathbf{r}_l) \times (\overline{\overline{\mathbf{G}}}_{E}(\mathbf{r}_l,\mathbf{\tilde{r}}) -  \overline{\overline{\mathbf{G}}}_{E}(\mathbf{r}_{h},\mathbf{\tilde{r}}))=0,
\end{equation}
\begin{equation}
\label{eqn:mbc}
\hat{\mathbf{n}}_l(\mathbf{r}_l) \times ( \varepsilon_l(\mathbf{r}_l) \overline{\overline{\mathbf{G}}}_{m}(\mathbf{r}_l,\mathbf{\tilde{r}})-\varepsilon_{h}(\mathbf{r}_h) \overline{\overline{\mathbf{G}}}_{m}(\mathbf{r}_{h},\mathbf{\tilde{r}})) = 0, 
\end{equation}
\begin{equation}
\label{eqn:Mbc}
\hat{\mathbf{n}}_l(\mathbf{r}_l) \times (\overline{\overline{\mathbf{G}}}_{M}(\mathbf{r}_l,\mathbf{\tilde{r}}) -  \overline{\overline{\mathbf{G}}}_{M}(\mathbf{r}_{h},\mathbf{\tilde{r}}))=0,
\end{equation}
\end{subequations}
where $\mathbf{r}_l$ and $\mathbf{r}_{h}$ are position vectors of points on either side of $S_l$ in volume $V_l$ and $V_{h}$ respectively ($\lvert \mathbf{r}_l - \mathbf{r}_h \rvert \rightarrow 0  $). In addition, the DGFs satisfy the following reciprocity relations: 
\begin{equation}
\label{eqn:erecip}
\mu(\mathbf{r})\overline{\overline{G}}^T_e(\mathbf{r},\mathbf{\tilde{r}}) = \mu(\mathbf{\tilde{r}}) \overline{\overline{G}}_e(\mathbf{\tilde{r}},\mathbf{r}),
\end{equation}
\begin{equation}
\label{eqn:mrecip}
\varepsilon(\mathbf{r})\overline{\overline{G}}^T_m(\mathbf{r},\mathbf{\tilde{r}}) = \varepsilon(\mathbf{\tilde{r}}) \overline{\overline{G}}_m(\mathbf{\tilde{r}},\mathbf{r}),
\end{equation}
\begin{equation}
\label{eqn:EMrecip}
\overline{\overline{G}}^T_E(\mathbf{r},\mathbf{\tilde{r}}) =  \overline{\overline{G}}_M(\mathbf{\tilde{r}},\mathbf{r}),
\end{equation}
where $ \overline{\overline{A}}^T $ is the transpose of $ \overline{\overline{A}} $. 

Radiative transfer can be determined from the Poynting vector, $ \mathbf{P}(\mathbf{\tilde{r}}) = \langle  \mathbf{E}(\mathbf{\tilde{r}},t) \times  \mathbf{H}(\mathbf{\tilde{r}},t) \rangle $, whose components are given by:
\begin{equation}
\label{eqn:poyntingvector} P_i(\mathbf{\tilde{r}}) = \epsilon_{ipq} \langle E_p(\mathbf{\tilde{r}},t) H_q(\mathbf{\tilde{r}},t) \rangle
\end{equation}
where $ \epsilon_{ipq} $ is the Levi-Civita symbol. 
To determine van der Waals pressure and radiative transfer, we need equal time correlations of various components of the electric and magnetic field vectors, such as $\langle E_p(\mathbf{\tilde{r}},t) E_q(\mathbf{\tilde{r}},t) \rangle$, $\langle H_p(\mathbf{\tilde{r}},t) H_q(\mathbf{\tilde{r}},t) \rangle$, and $\langle E_p(\mathbf{\tilde{r}},t) H_q(\mathbf{\tilde{r}},t) \rangle$. van der Waals pressure in vacuum can be determined from the Maxwell stress tensor, $ \overline{\overline{\sigma}}=\overline{\overline{\sigma}}^e+\overline{\overline{\sigma}}^m $, where $ \overline{\overline{\sigma}}^e $ and $ \overline{\overline{\sigma}}^m $ are the electric and magnetic field contributions respectively. $ \overline{\overline{\sigma}}^e $ and $ \overline{\overline{\sigma}}^m $ are given by:
\begin{equation}
\label{eqn:efieldstresstensor}
\overline{\overline{\sigma}}^e(\mathbf{\tilde{r}}) = \varepsilon_o \left[ \langle \mathbf{E}(\mathbf{\tilde{r}},t) \mathbf{E}(\mathbf{\tilde{r}},t) \rangle -\frac{1}{2} \overline{\overline{I}} \langle \mathbf{E}^2(\mathbf{\tilde{r}},t) \rangle \right]
\end{equation}
\begin{equation}
\label{eqn:mfieldstresstensor}
\overline{\overline{\sigma}}^m(\mathbf{\tilde{r}}) = \mu_o \left[ \langle \mathbf{H}(\mathbf{\tilde{r}},t) \mathbf{H}(\mathbf{\tilde{r}},t) \rangle -\frac{1}{2} \overline{\overline{I}} \langle \mathbf{H}^2(\mathbf{\tilde{r}},t) \rangle \right],
\end{equation}
where $ \langle \mathbf{E}(\mathbf{\tilde{r}},t) \mathbf{E}(\mathbf{\tilde{r}},t) \rangle $ and $ \langle \mathbf{H}(\mathbf{\tilde{r}},t) \mathbf{H}(\mathbf{\tilde{r}},t) \rangle $ are matrices whose components are $ \langle E_p(\mathbf{\tilde{r}},t) E_q(\mathbf{\tilde{r}},t) \rangle $ and $ \langle H_p(\mathbf{\tilde{r}},t) H_q(\mathbf{\tilde{r}},t) \rangle $ respectively, $ p,q=1,2,3 $, and $ \overline{\overline{I}} $ is the identity matrix.

 Since the fields are assumed to be stationary, $ \langle \mathbf{E}(\mathbf{\tilde{r}},t) \mathbf{E}(\mathbf{\tilde{r}},t) \rangle $, $ \langle \mathbf{H}(\mathbf{\tilde{r}},t) \mathbf{H}(\mathbf{\tilde{r}},t) \rangle $, and $ \langle \mathbf{E}(\mathbf{\tilde{r}},t) \mathbf{H}(\mathbf{\tilde{r}},t) \rangle $ are independent of time \cite{mandel1995optical}. 
The equal time correlation functions are related to the cross-spectral densities by:
\begin{equation}
\label{eqn:epeqtime}
\begin{split}
\langle E_p(\mathbf{\tilde{r}},t) E_q(\mathbf{\tilde{r}},t)  & \rangle  =\int\limits_0^{\infty} \frac{d\omega}{2\pi} \langle  E_p(\mathbf{\tilde{r}}) E^*_q(\mathbf{\tilde{r}}) \rangle_s  \\
=& \int\limits_0^{\infty} \frac{d\omega}{2\pi}  \langle  E_p(\mathbf{\tilde{r}}) E^*_q(\mathbf{\tilde{r}}) +  E_p^*(\mathbf{\tilde{r}}) E_q(\mathbf{\tilde{r}}) \rangle
\end{split}
\end{equation}
\begin{equation}
\label{eqn:hphqtime}
\begin{split}
\langle H_p(\mathbf{\tilde{r}},t) H_q(\mathbf{\tilde{r}},t)  & \rangle=\int\limits_0^{\infty} \frac{d\omega}{2\pi} \langle  H_p(\mathbf{\tilde{r}}) H^*_q(\mathbf{\tilde{r}}) \rangle_s \\
=& \int\limits_0^{\infty} \frac{d\omega}{2\pi}  \langle H_p(\mathbf{\tilde{r}}) H^*_q(\mathbf{\tilde{r}}) + H_p^*(\mathbf{\tilde{r}}) H_q(\mathbf{\tilde{r}}) \rangle
\end{split}
\end{equation}
\begin{equation}
\label{eqn:ephqtime}
\begin{split}
\langle E_p(\mathbf{\tilde{r}},t) H_q(\mathbf{\tilde{r}},t)   & \rangle=\int\limits_0^{\infty} \frac{d\omega}{2\pi} \langle  E_p(\mathbf{\tilde{r}}) H^*_q(\mathbf{\tilde{r}}) \rangle_s\\
=& \int\limits_0^{\infty} \frac{d\omega}{2\pi}  \langle E_p(\mathbf{\tilde{r}}) H^*_q(\mathbf{\tilde{r}}) + E_p^*(\mathbf{\tilde{r}}) H_q(\mathbf{\tilde{r}}) \rangle
\end{split}
\end{equation}
where the subscript $ s $ in Eq. \ref{eqn:epeqtime} - Eq. \ref{eqn:ephqtime} implies a symmetric sum. 
Using Eq. \ref{eqn:fdresults} and Eq. \ref{eqn:fieldrelations}, we can express the cross-spectral densities of the components of the electric and magnetic field at $\mathbf{\tilde{r}} \in V_{h}$ as:
\begin{equation}
\label{eqn:eecorr} \langle  \mathbf{E}(\mathbf{\tilde{r}})   \mathbf{E}^*(\mathbf{\tilde{r}}) \rangle_{s}  = 2\omega \mu_o \left[ \sum\limits_{l=1}^N \Theta_l \overline{\overline{\mathcal{E}}}^{(l)}  (\mathbf{\tilde{r}})  + \Theta_h \overline{\overline{\mathcal{E}}}^{(h)}  (\mathbf{\tilde{r}})\right] 
\end{equation}
\begin{equation}
\label{eqn:hhcorr} \langle  \mathbf{H}(\mathbf{\tilde{r}})   \mathbf{H}^*(\mathbf{\tilde{r}}) \rangle _{s} =  2\omega \varepsilon_o \left[\sum\limits_{l=1}^N \Theta_l \overline{\overline{\mathcal{H}}}^{(l)}(\mathbf{\tilde{r}}) + \Theta_h \overline{\overline{\mathcal{H}}}^{(h)} (\mathbf{\tilde{r}}) \right] 
\end{equation}
\begin{equation}
\label{eqn:ehcorr}\langle  \mathbf{E}(\mathbf{\tilde{r}})   \mathbf{H}^*(\mathbf{\tilde{r}}) \rangle  =  \sum\limits_{l=1}^N \Theta_l \overline{\overline{\mathcal{X}}}^{(l)}\left(\mathbf{\tilde{r}}\right) + \Theta_h \overline{\overline{\mathcal{X}}}^{(h)}\left(\mathbf{\tilde{r}}\right)
\end{equation}
where $ \Theta_l = \Theta(\omega,T_l) $, and $ \langle \mathbf{E}(\mathbf{\tilde{r}})   \mathbf{E}^*(\mathbf{\tilde{r}}) \rangle_s $, $ \langle \mathbf{H}(\mathbf{\tilde{r}})   \mathbf{H}^*(\mathbf{\tilde{r}}) \rangle_s $, and $ \langle \mathbf{E}(\mathbf{\tilde{r}})   \mathbf{H}^*(\mathbf{\tilde{r}}) \rangle $ are matrices whose components are $ \langle E_p(\mathbf{\tilde{r}}) E^*_q(\mathbf{\tilde{r}}) \rangle_{s} $, $ \langle H_p(\mathbf{\tilde{r}}) H^*_q(\mathbf{\tilde{r}}) \rangle_{s} $, and $ \langle E_p(\mathbf{\tilde{r}}) H^*_q(\mathbf{\tilde{r}}) \rangle $ respectively. Even though components of $ \langle  \mathbf{E}   \mathbf{H}^* \rangle_s $ are necessary to compute radiative transfer, we persist with $ \langle  \mathbf{E}   \mathbf{H}^* \rangle $. The reason for computing $ \langle  \mathbf{E}   \mathbf{H}^* \rangle $ as opposed to $ \langle  \mathbf{E}   \mathbf{H}^* \rangle_s $ will be clarified in Sec. \ref{sec:fesif}. 

The matrices $\overline{\overline{\mathcal{E}}}^{(l)}\left(\mathbf{\tilde{r}}\right)$, $\overline{\overline{\mathcal{H}}}^{(l)}\left(\mathbf{\tilde{r}}\right)$, and $\overline{\overline{\mathcal{X}}}^{(l)}\left(\mathbf{\tilde{r}}\right)$ are contributions to $ \langle \mathbf{E}(\mathbf{\tilde{r}})    \mathbf{E}^*(\mathbf{\tilde{r}})  \rangle_s $, $ \langle \mathbf{H}(\mathbf{\tilde{r}})    \mathbf{H}^*(\mathbf{\tilde{r}})  \rangle_s $, and $ \langle \mathbf{E}(\mathbf{\tilde{r}})    \mathbf{H}^*(\mathbf{\tilde{r}})  \rangle $ ($ l=1,2,\cdots,N,h $) from sources in volume $ V_l $. 
For $ l \in \{1,2,\cdots,N\}, $ $\overline{\overline{\mathcal{E}}}^{(l)}\left(\mathbf{\tilde{r}}\right)$, $\overline{\overline{\mathcal{H}}}^{(l)}\left(\mathbf{\tilde{r}}\right)$, and $\overline{\overline{\mathcal{X}}}^{(l)}\left(\mathbf{\tilde{r}}\right)$ are given by: 
\begin{subequations}
\begin{equation}
\label{eqn:eqep*}
\begin{split}
 \overline{\overline{\mathcal{E}}}^{(l)} (\mathbf{\tilde{r}})   =   2\Re  \int\limits_{V_l}  &  d\mathbf{r} \Big[  \mu''(\mathbf{r})   \overline{\overline{G}}^T_{E}(\mathbf{r}, \mathbf{\tilde{r}}) \cdot \overline{\overline{G}}^*_{E}(\mathbf{r}, \mathbf{\tilde{r}})   + \\
& \varepsilon''(\mathbf{r}) |\mu(\mathbf{r})|^2\frac{\omega^2}{c^2}  \overline{\overline{G}}^T_{e}(\mathbf{r}, \mathbf{\tilde{r}})  \cdot \overline{\overline{G}}^*_{e}(\mathbf{r}, \mathbf{\tilde{r}})  \Big] 
 \end{split}
 \end{equation}
 \begin{equation}
 \label{eqn:hqhp*}
\begin{split} 
  \overline{\overline{\mathcal{H}}}^{(l)} \left(\mathbf{\tilde{r}}\right)  = 2 \Re  & \int\limits_{V_l}  d\mathbf{r}  \Big[  \varepsilon''(\mathbf{r})  \overline{\overline{G}}^T_{M}(\mathbf{r}, \mathbf{\tilde{r}}) \cdot \overline{\overline{G}}^*_{M}(\mathbf{r}, \mathbf{\tilde{r}})   +  \\ 
& \mu''(\mathbf{r}) |\varepsilon(\mathbf{r})|^2\frac{\omega^2}{c^2}  \overline{\overline{G}}^T_{m}(\mathbf{r}, \mathbf{\tilde{r}})  \cdot \overline{\overline{G}}^*_{m}(\mathbf{r}, \mathbf{\tilde{r}}) 
 \Big] 
 \end{split}
 \end{equation}
 \begin{equation}
 \label{eqn:eqhp*}
 \begin{split}
\overline{\overline{\mathcal{X}}}^{(l)} \left(\mathbf{\tilde{r}}\right)  =   i 2\frac{\omega^2}{c^2}   & \int\limits_{V_l}  \Big[ \varepsilon''(\mathbf{r}) \mu(\mathbf{r})  \overline{\overline{G}}^T_{e}(\mathbf{r}, \mathbf{\tilde{r}})  \cdot \overline{\overline{G}}^*_{M}(\mathbf{r}, \mathbf{\tilde{r}})   + \\
&  \mu''(\mathbf{r})  \varepsilon^*(\mathbf{r})  \overline{\overline{G}}^T_{E}(\mathbf{r}, \mathbf{\tilde{r}}) \cdot \overline{\overline{G}}^*_{m}(\mathbf{r}, \mathbf{\tilde{r}}) \Big] d\mathbf{r}
 \end{split}
\end{equation}
\end{subequations}
The expressions for $\overline{\overline{\mathcal{E}}}^{(h)}\left(\mathbf{\tilde{r}}\right)$, $\overline{\overline{\mathcal{H}}}^{(h)}\left(\mathbf{\tilde{r}}\right)$, and $\overline{\overline{\mathcal{X}}}^{(h)}\left(\mathbf{\tilde{r}}\right)$ have to be modified to take into account the singularity of the DGFs in the integrals in Eq. \ref{eqn:eqep*singularity}-\ref{eqn:eqhp*singularity} as $ \lvert \mathbf{r} - \mathbf{\tilde{r}} \rvert \rightarrow 0 $. The modified expressions for $\overline{\overline{\mathcal{E}}}^{(h)}\left(\mathbf{\tilde{r}}\right)$, $\overline{\overline{\mathcal{H}}}^{(h)}\left(\mathbf{\tilde{r}}\right)$, and $\overline{\overline{\mathcal{X}}}^{(h)}\left(\mathbf{\tilde{r}}\right)$ are:
\begin{subequations}
\begin{equation}
\label{eqn:eqep*singularity}
\begin{split}
 \overline{\overline{\mathcal{E}}}^{(h)} & (\mathbf{\tilde{r}})    =   \lim_{V_{\delta} \rightarrow 0}  2\Re   \int\limits_{V_h}   \Big[ \mu''(\mathbf{r})   \overline{\overline{G}}^T_{E}(\mathbf{r}, \mathbf{\tilde{r}}) \cdot \overline{\overline{G}}^*_{E}(\mathbf{r}, \mathbf{\tilde{r}})   \\
+& \varepsilon''(\mathbf{r})  |\mu(\mathbf{r})|^2\frac{\omega^2}{c^2}  \overline{\overline{G}}^T_{e}(\mathbf{r}, \mathbf{\tilde{r}})  \cdot \overline{\overline{G}}^*_{e}(\mathbf{r}, \mathbf{\tilde{r}})  \Big] d\mathbf{r}  \\
+ & \varepsilon_h'' 2 \Re \Big( \frac{\mu_h}{\varepsilon_h^{*}} \overline{\overline{L}}   \cdot  \overline{\overline{G}}_e^{(sc)}(\mathbf{\tilde{r}},\mathbf{\tilde{r}})   +       \frac{\mu_h}{\varepsilon^*_h}  \overline{\overline{G}}_e^{(sc)T}(\mathbf{\tilde{r}},\mathbf{\tilde{r}})   \cdot  \overline{\overline{L}}  \Big)
 \end{split}
 \end{equation}
  \begin{equation}
 \label{eqn:hqhp*singularity}
\begin{split} 
  \overline{\overline{\mathcal{H}}}^{(h)} & \left(\mathbf{\tilde{r}}\right)  =  \lim_{V_{\delta} \rightarrow 0}  2\Re    \int\limits_{V_h}   \Big[  \varepsilon''(\mathbf{r})  \overline{\overline{G}}^T_{M}(\mathbf{r}, \mathbf{\tilde{r}}) \cdot \overline{\overline{G}}^*_{M}(\mathbf{r}, \mathbf{\tilde{r}})  \\
+& \mu''(\mathbf{r})  |\varepsilon(\mathbf{r}) |^2  \frac{\omega^2}{c^2}  \overline{\overline{G}}^T_{m}(\mathbf{r}, \mathbf{\tilde{r}})  \cdot \overline{\overline{G}}^*_{m}(\mathbf{r}, \mathbf{\tilde{r}}) 
 \Big] d\mathbf{r}  \\
+ & \mu''_h 2 \Re \Big( \frac{\varepsilon_h}{\mu^{*}_h}  \overline{\overline{L}}  \cdot    \overline{\overline{G}}_m^{(sc)}(\mathbf{\tilde{r}},\mathbf{\tilde{r}})     + \frac{\varepsilon_h}{\mu^*_h}    \overline{\overline{G}}_m^{(sc)T}(\mathbf{\tilde{r}},\mathbf{\tilde{r}})  \cdot \overline{\overline{L}} \Big) 
 \end{split}
 \end{equation}
 \begin{equation}
 \label{eqn:eqhp*singularity}
 \begin{split}
\overline{\overline{\mathcal{X}}}^{(h)} & \left(\mathbf{\tilde{r}}\right)  =   i  2   \frac{\omega^2}{c^2}    \lim_{V_{\delta} \rightarrow 0} \int\limits_{V_l}  \Big[ \varepsilon''(\mathbf{r}) \mu(\mathbf{r})  \overline{\overline{G}}^T_{e}(\mathbf{r}, \mathbf{\tilde{r}})  \cdot \overline{\overline{G}}^*_{M}(\mathbf{r}, \mathbf{\tilde{r}})  \\
 &  + \mu''(\mathbf{r})  \varepsilon^*(\mathbf{r})  \overline{\overline{G}}^T_{E}(\mathbf{r}, \mathbf{\tilde{r}}) \cdot \overline{\overline{G}}^*_{m}(\mathbf{r}, \mathbf{\tilde{r}}) \Big] d\mathbf{r}  \\
& - i 2 \Big( \frac{\varepsilon''_h}{\varepsilon_h} \overline{\overline{L}} \cdot   \overline{\overline{G}}_M^{(sc)*}(\mathbf{\tilde{r}},\mathbf{\tilde{r}})     +  \frac{\mu''_h}{\mu^*_h}   \overline{\overline{G}}_E^{(sc)T}(\mathbf{\tilde{r}},\mathbf{\tilde{r}})  \cdot  \overline{\overline{L}}  \Big)   
 \end{split}
\end{equation}
\end{subequations}
where 
$V_{\delta}$ is volume of infinitesimal radius surrounding $\mathbf{\tilde{r}}$, $\overline{\overline{L}}$ is a shape dependent dyad \cite{Yaghjian80}, and $\overline{\overline{G}}_e(\mathbf{r},\mathbf{\tilde{r}}) =\overline{\overline{G}}_o (\mathbf{r},\mathbf{\tilde{r}})+\overline{\overline{G}}^{(sc)}_e(\mathbf{r},\mathbf{\tilde{r}})$, $\overline{\overline{G}}_m(\mathbf{r},\mathbf{\tilde{r}}) =\overline{\overline{G}}_o (\mathbf{r},\mathbf{\tilde{r}})+\overline{\overline{G}}^{(sc)}_m(\mathbf{r},\mathbf{\tilde{r}})$, $\overline{\overline{G}}_E(\mathbf{r},\mathbf{\tilde{r}}) =\nabla \times \overline{\overline{G}}_o (\mathbf{r},\mathbf{\tilde{r}})+\overline{\overline{G}}^{(sc)}_E(\mathbf{r},\mathbf{\tilde{r}})$, and $\overline{\overline{G}}_M(\mathbf{r},\mathbf{\tilde{r}}) =\nabla \times \overline{\overline{G}}_o (\mathbf{r},\mathbf{\tilde{r}})+\overline{\overline{G}}^{(sc)}_M(\mathbf{r},\mathbf{\tilde{r}})$. $\overline{\overline{G}}_o (\mathbf{r},\mathbf{\tilde{r}})$ is the DGF when no scatterers are present, and $\overline{\overline{G}}^{(sc)}(\mathbf{r},\mathbf{\tilde{r}})$ is the contribution from presence of scatterers. $\overline{\overline{G}}^{(sc)}(\mathbf{r},\mathbf{\tilde{r}})$ is always finite. The volume integrals in Eq. \ref{eqn:eqep*singularity}-Eq. \ref{eqn:eqhp*singularity} are finite even though $ \varepsilon_h = \mu_h = 0 $ because of the singularity in the DGFs. The last line of Eq. \ref{eqn:eqep*singularity}-Eq. \ref{eqn:eqhp*singularity} are identically equal to zero for non-absorbing materials, including vacuum. However, we choose to retain them since they are essential for calculations of cross-spectral densities in absorbing media. We have neglected terms independent of the configuration of the scatterers. These terms are infinite because of the assumption that the thermal sources at any two locations are uncorrelated. They can be made finite by eliminating the local assumption in Eq. \ref{eqn:fdresults} but that will not affect the calculations of forces or heat transfer except at gaps smaller than the correlation length. Usually, the correlation length is of the order of the atomic spacing in dielectrics or the electron mean free path in metals. More detailed discussion of the singularity in DGFs and calculation of cross-spectral densities is given in Ref. \cite{narayanaswamy2010a}. 

\section{\label{sec:fesif}Surface integral dyadic Green's function formalism}
While the volume integrals in Eq. \ref{eqn:eqep*} - Eq. \ref{eqn:eqhp*singularity} can in principle be used to compute forces and radiative heat transfer, they are undesirable for the following reasons: 
(1) Evaluating classical radiative transfer between two objects requires the computation of the view factor between them. But for objects with simple geometries, computation of the view factor between two objects requires, in general, not the evaluation of a volume integral but the evaluation of a double integral over the surfaces of the two objects,  
(2) the expressions in Eq. \ref{eqn:eqep*} - Eq. \ref{eqn:eqhp*singularity} do not reflect the different reciprocity relations and boundary conditions satisfied by the DGFs, and (3) evaluation of volume integrals are computationally more expensive than that of surface integrals. These undesirable features can be overcome by converting Eq. \ref{eqn:eqep*} - Eq. \ref{eqn:eqhp*singularity} into appropriate surface integrals using Green's theorems for dyadic functions \cite{Tai93}. The surface integral representations for the cross-spectral densities are as follows: 
\begin{subequations}
\begin{equation}
\label{eqn:eqep*surf}
  \overline{\overline{\mathcal{E}}}^{(l)} (\mathbf{\tilde{r}})  =  2\Im \oint\limits_{S_l}  \big[ \mu(\mathbf{r}) \overline{\overline{G}}^T_e (\mathbf{r}, \mathbf{\tilde{r}})\big]^* \cdot [\mathbf{n}_{l}(\mathbf{r}) \times \overline{\overline{G}}_E (\mathbf{r}, \mathbf{\tilde{r}} )] d\mathbf{r} 
 \end{equation}
\begin{equation}
\label{eqn:hqhp*surf}
\overline{\overline{\mathcal{H}}}^{(l)} (\mathbf{\tilde{r}})  =  2\Im \oint\limits_{S_l}  \big[\varepsilon(\mathbf{r}) \overline{\overline{G}}^{T}_m (\mathbf{r}, \mathbf{\tilde{r}}) \big]^* \cdot [\mathbf{n}_{l}(\mathbf{r}) \times \overline{\overline{G}}_M (\mathbf{r}, \mathbf{\tilde{r}}) ]  d\mathbf{r}
 \end{equation}
 \begin{equation}
 \label{eqn:eqhp*surf}
 \begin{split}
 \overline{\overline{\mathcal{X}}}^{(l)} (\mathbf{\tilde{r}})   & = - \oint\limits_{S_l}   \Big[ 
\overline{\overline{G}}^T_E \left(\mathbf{r}, \mathbf{\tilde{r}}\right) . (\mathbf{n}_{l}(\mathbf{r}) \times \overline{\overline{G}}_M^* (\mathbf{r}, \mathbf{\tilde{r}}) )   + \\
& \frac{\omega^2}{c^2}  \mu(\mathbf{r})\overline{\overline{G}}^T_e  (\mathbf{r}, \mathbf{\tilde{r}}) . (\mathbf{n}_{l}(\mathbf{r}) \times \varepsilon(\mathbf{r}) \overline{\overline{G}}_m (\mathbf{r}, \mathbf{\tilde{r}}) )^* 
\Big] d\mathbf{r}
 \end{split}
 \end{equation}
 \begin{equation}
  \label{eqn:eqep*surfhost}
 \begin{split}
 \overline{\overline{\mathcal{E}}}^{(h)} (\mathbf{\tilde{r}})   =  -2 \sum\limits_{l=1}^N & \Im \oint\limits_{S_l} \mu^*(\mathbf{r}) \overline{\overline{G}}^{*T}_e (\mathbf{r}, \mathbf{\tilde{r}}) \cdot (\mathbf{n}_{l}(\mathbf{r}) \times \overline{\overline{G}}_E (\mathbf{r}, \mathbf{\tilde{r}} )) d\mathbf{r}   \\
 + & \Im (2\mu_h \overline{\overline{G}}_e ( \mathbf{\tilde{r}},\mathbf{\tilde{r}} ) )
 \end{split}
 \end{equation}
 \begin{equation}
 \label{eqn:hqhp*surfhost}
 \begin{split}
\overline{\overline{\mathcal{H}}}^{(h)} (\mathbf{\tilde{r}})   =  -2\sum\limits_{l=1}^N &\Im  \oint\limits_{S_l} \varepsilon^*(\mathbf{r}) \overline{\overline{G}}^{*T}_m (\mathbf{r}, \mathbf{\tilde{r}}) \cdot (\mathbf{n}_{l}(\mathbf{r}) \times \overline{\overline{G}}_M (\mathbf{r}, \mathbf{\tilde{r}}) )  d\mathbf{r}  \\
+ &  \Im (2\varepsilon_h \overline{\overline{G}}_m ( \mathbf{\tilde{r}},\mathbf{\tilde{r}}) )
\end{split}
 \end{equation}
 \begin{equation}
 \label{eqn:eqhp*surfhost}
 \begin{split}
\overline{\overline{\mathcal{X}}}^{(h)} (\mathbf{\tilde{r}})  &  =   \sum\limits_{l=1}^N \oint\limits_{S_l}  \Big[ 
\overline{\overline{G}}^T_E (\mathbf{r}, \mathbf{\tilde{r}}) \cdot (\mathbf{n}_{l}(\mathbf{r}) \times \overline{\overline{G}}_M^* (\mathbf{r}, \mathbf{\tilde{r}}) )   +  \\
& \frac{\omega^2}{c^2}  \mu(\mathbf{r})  \overline{\overline{G}}^T_e  (\mathbf{r}, \mathbf{\tilde{r}}) \cdot  (\mathbf{n}_{l}(\mathbf{r}) \times \varepsilon(\mathbf{r}) \overline{\overline{G}}_m (\mathbf{r}, \mathbf{\tilde{r}}) )^* 
\Big]d\mathbf{r}  \\
& - i2\Im (\overline{\overline{G}}_M(\mathbf{\tilde{r}}, \mathbf{\tilde{r}}) )
\end{split}
 \end{equation}
\end{subequations}

If the host medium is dissipative, the functions $ \Im (2\mu_h \overline{\overline{G}}_e ( \mathbf{\tilde{r}},\mathbf{\tilde{r}} ) ) $, $ \Im (2\varepsilon_h \overline{\overline{G}}_m ( \mathbf{\tilde{r}},\mathbf{\tilde{r}}) ) $, and $ \Im (\overline{\overline{G}}_M (\mathbf{\tilde{r}}, \mathbf{\tilde{r}}) ) $ in Eq. \ref{eqn:eqep*surfhost}, Eq. \ref{eqn:hqhp*surfhost}, and Eq. \ref{eqn:eqhp*surfhost} respectively should be replaced by $ \Im (2\mu_h \overline{\overline{G}}_e^{(sc)} ( \mathbf{\tilde{r}},\mathbf{\tilde{r}} ) ) $, $ \Im (2\varepsilon_h \overline{\overline{G}}_m^{(sc)} ( \mathbf{\tilde{r}},\mathbf{\tilde{r}}) ) $, and $ \Im (\overline{\overline{G}}_M^{(sc)} (\mathbf{\tilde{r}}, \mathbf{\tilde{r}}) ) $. Using Eq. \ref{eqn:eqep*surf}-\ref{eqn:eqhp*surfhost}, Eq. \ref{eqn:eecorr}-\ref{eqn:ehcorr} can be re-written as:
\begin{equation}
\label{eqn:eecorrsurf}
\begin{split}
\langle  \mathbf{E}   \mathbf{E}^* \rangle_{s}  = & 2\omega \mu_o \sum\limits_{l=1}^N \left(\Theta_l-\Theta_h\right) \overline{\overline{\mathcal{E}}}^{(l)}  (\mathbf{\tilde{r}})  +  2\omega \mu_o \Theta_h \Im (2\mu_h \overline{\overline{G}}_e ( \mathbf{\tilde{r}},\mathbf{\tilde{r}}) ) 
\end{split}
\end{equation}
\begin{equation}
\label{eqn:hhcorrsurf}
\begin{split}
 \langle  \mathbf{H}   \mathbf{H}^* \rangle _{s} = &  2\omega \varepsilon_o \sum\limits_{l=1}^N \left(\Theta_l-\Theta_h\right) \overline{\overline{\mathcal{H}}}^{(l)}  (\mathbf{\tilde{r}})  + 2\omega \varepsilon_o \Theta_h \Im (2\varepsilon_h \overline{\overline{G}}_m ( \mathbf{\tilde{r}},\mathbf{\tilde{r}}) ) 
 \end{split}
 \end{equation}
\begin{equation} 
\label{eqn:ehcorrsurf}
 \langle  \mathbf{E}   \mathbf{H}^* \rangle  =   \sum\limits_{l=1}^N (\Theta_l - \Theta_h) \overline{\overline{\mathcal{X}}}^{(l)}(\mathbf{\tilde{r}}) -i 2 \Theta_h \Im \overline{\overline{G}}_M(\mathbf{\tilde{r}}, \mathbf{\tilde{r}}) 
\end{equation}
The terms $ \Im (2\mu_h \overline{\overline{G}}_e ( \mathbf{\tilde{r}},\mathbf{\tilde{r}}) ) $ and $ \Im (2\varepsilon_h \overline{\overline{G}}_m ( \mathbf{\tilde{r}},\mathbf{\tilde{r}}) ) $ are thermal equilibrium contributions to $  \overline{\overline{\mathcal{E}}}^{(h)} (\mathbf{\tilde{r}}) $ and $  \overline{\overline{\mathcal{H}}}^{(h)} (\mathbf{\tilde{r}}) $ respectively. They give rise to the van der Waals stresses as predicted by Lifshitz theory when $\varepsilon_h = \mu_h = 1$.
The reason for persisting with $ \langle  \mathbf{E}   \mathbf{H}^* \rangle $, as opposed to $ \langle  \mathbf{E}   \mathbf{H}^* \rangle_s =  \mathbf{E}   \mathbf{H}^* +  \mathbf{E}^* \mathbf{H}$, is to show that there is indeed an equilibrium contribution to $ \langle  \mathbf{E}   \mathbf{H}^* \rangle $. However, when we compute $ \langle  \mathbf{E}   \mathbf{H}^* \rangle_s $, the equilibrium contribution vanishes since radiative energy transfer between two objects at the same temperature must be zero.

Radiative transfer between two objects is discussed further in Sec. \ref{sec:generalizedtransmissivity}.  Before proceeding to Sec. \ref{sec:generalizedtransmissivity}, we wish to remark on the form of the non-equilibrium contributions in Eq. \ref{eqn:eqep*surf}-\ref{eqn:eqhp*surfhost}. Using the property that $\overline{\overline{A}} = \mathbf{n} \left(\mathbf{n} \cdot \overline{\overline{A}}\right) - \mathbf{n} \times \mathbf{n} \times \overline{\overline{A}}$, we see that all the surface integrals in Eq. \ref{eqn:eqep*surf}-\ref{eqn:eqhp*surfhost} feature only tangential components of the dyadic Green's functions that are continuous across an interface between two materials. 



\section{\label{sec:generalizedtransmissivity}Generalized transmissivity for radiative energy transfer}

 The steady state radiative heat transfer from object $ 1 $ to object $ 2 $ in Fig. \ref{fig:generalinteraction}, $ Q_{1 \rightarrow 2} $, is given by:
\begin{equation}
\label{eqn:Q12surfintegral}
Q_{1 \rightarrow 2} = -\oint\limits_{S_2} \mathbf{P}^{(1)}(\mathbf{\tilde{r}}) \cdot \mathbf{n}_2(\mathbf{\tilde{r}}) d\mathbf{\tilde{r}} 
\end{equation}
where $\mathbf{P}^{(1)}(\mathbf{\tilde{r}})$ is the Poynting vector at $\mathbf{\tilde{r}} \in S_2$ due to thermally fluctuating sources within $V_1$. The ``$ - $'' sign in front of the surface integral is because $ \mathbf{n}_2(\mathbf{\tilde{r}}) $ is the outward pointing normal on the surface $ S_2 $. The net heat transfer between objects $ 1 $ and $ 2 $ is given by $Q_{1,2} =  Q_{1 \rightarrow 2} - Q_{2 \rightarrow 1} $. The components of $\mathbf{P}^{(1)}(\mathbf{\tilde{r}})$ are given by:
\begin{equation}
\label{eqn:poyntingcomponent}
P^{(1)}_i(\mathbf{\tilde{r}})=\epsilon_{iqp} \langle E_q(\mathbf{\tilde{r}},t) H_p(\mathbf{\tilde{r}},t) \rangle^{(1)}, 
\end{equation}
where $ \langle E_q(\mathbf{\tilde{r}},t) H_p(\mathbf{\tilde{r}},t) \rangle^{(1)} $ is the contribution to $ \langle E_q(\mathbf{\tilde{r}},t) H_p(\mathbf{\tilde{r}},t) \rangle $ from sources within $ V_1 $. The object indices $ 1 $ and $ 2 $ can be replaced by any $ m,n \in \{1,2,\cdots,N\} $. From Sec. \ref{sec:fegff}, we know that $\displaystyle \langle E_q(\mathbf{\tilde{r}},t) H_p(\mathbf{\tilde{r}},t) \rangle^{(1)} =  \int\limits_0^{\infty} \frac{d\omega}{2\pi}  \langle E_q(\mathbf{\tilde{r}},\omega) H_p^*(\mathbf{\tilde{r}},\omega) \rangle_s^{(1)} $. Using Eq. \ref{eqn:ehcorr}, Eq. \ref{eqn:ehcorrsurf}, Eq. \ref{eqn:eqhp*surf}, and Eq. \ref{eqn:poyntingcomponent}, Eq. \ref{eqn:Q12surfintegral} for $ Q_{1 \rightarrow 2} $ can be re-written as:
\begin{equation}
\label{eqn:T12definition}
\begin{split}
 Q_{1 \rightarrow 2} = & -\int\limits_{0}^{\infty} \frac{d\omega}{2\pi} \Theta(\omega,T_1) \oint\limits_{S_2} d\mathbf{\tilde{r}} n_{2i}(\mathbf{\tilde{r}}) \epsilon_{ipq} 2\Re\mathcal{X}^{(1)}_{pq}(\mathbf{\tilde{r}}) \\
 = & \int\limits_{0}^{\infty} \frac{d\omega}{2\pi} \Theta(\omega,T_1)  T^{e}_{1\rightarrow 2}(\omega) \\
 \Rightarrow T^{e}_{1\rightarrow 2}&(\omega)  =  -\oint\limits_{S_2} d\mathbf{\tilde{r}} n_{2i}(\mathbf{\tilde{r}}) \epsilon_{ipq} 2\Re\mathcal{X}^{(1)}_{pq}(\mathbf{\tilde{r}}) \\
 & = -\oint\limits_{S_2} d\mathbf{\tilde{r}} \epsilon_{qip} n_{2i}(\mathbf{\tilde{r}}) 2\Re\mathcal{X}^{(1)}_{pq}(\mathbf{\tilde{r}}) \\
 & = - 2 \Re\oint\limits_{S_2} d\mathbf{\tilde{r}} \left(\mathbf{\hat{n}}_{2}(\mathbf{\tilde{r}}) \times \overline{\overline{\mathcal{X}}}^{(1)}(\mathbf{\tilde{r}}) \right)_{qq}\\
 & = - 2 \Re Tr \oint\limits_{S_2} d\mathbf{\tilde{r}} \left( \mathbf{\hat{n}}_{2}(\mathbf{\tilde{r}}) \times \overline{\overline{\mathcal{X}}}^{(1)}(\mathbf{\tilde{r}})  \right),
 \end{split}
\end{equation}
where $T^{e}_{1 \rightarrow 2}(\omega)$ is a generalized transmissivity for radiative energy transport between objects 1 and 2, and $Tr(\overline{\overline{A}}) = \sum\limits_{p=1}^{3} A_{pp}$. The superscript $ e $ in $T^{e}_{1 \rightarrow 2}(\omega)$ stands for ``energy.'' Substituting the expression for $ \overline{\overline{\mathcal{X}}}^{(1)} $  from Eq. \ref{eqn:eqhp*surf} in the last line of Eq. \ref{eqn:T12definition}, 
 $T^{e}_{1 \rightarrow 2}(\omega)$ can be shown to be:
\begin{equation}
\label{eqn:generalizedtransmissivity1}
\begin{split}
T^{e}_{1 \rightarrow 2}(\omega) 
& =  2\Re Tr   \oint\limits_{S_1} d\mathbf{r} \oint\limits_{S_2}  d\mathbf{\tilde{r}} \bigg[  \frac{\omega^2}{c^2}  \times \\
& 
  [\mathbf{\hat{n}}_2(\mathbf{\tilde{r}}) \times \mu_2 \overline{\overline{G}}_e(\mathbf{\tilde{r}},\mathbf{r})]  \cdot 
   [\mathbf{\hat{n}}_1(\mathbf{r}) \times \varepsilon_1^* \overline{\overline{G}}_m^*(\mathbf{r},\mathbf{\tilde{r}})]  \\
     & + [\mathbf{\hat{n}}_2(\mathbf{\tilde{r}})  \times  \overline{\overline{G}}_E(\mathbf{\tilde{r}},\mathbf{r})] \cdot 
  [\mathbf{\hat{n}}_1(\mathbf{r}) \times \overline{\overline{G}}^*_E(\mathbf{r},\mathbf{\tilde{r}})]  \bigg] 
\end{split}
\end{equation}
 where  $\mathbf{\hat{n}}_1(\mathbf{r})$ is the outward pointing normal on the surface   $S_1$, as shown in Fig. \ref{fig:generalinteraction}. For any two vectors $ \mathbf{a}, \mathbf{b} $ and dyads $ \overline{\overline{\mathbf{A}}}, \overline{\overline{\mathbf{B}}} $, the following  property can be shown to be true: $ Tr\big\lbrace \big(\mathbf{a} \times \overline{\overline{\mathbf{A}}}\big) \cdot \big(\mathbf{b} \times \overline{\overline{\mathbf{B}}} \big) \big\rbrace  = Tr\big\lbrace \big(\mathbf{b} \times \overline{\overline{\mathbf{A}}}^{T}\big) \cdot \big(\mathbf{a} \times \overline{\overline{\mathbf{B}}}^T \big) \big\rbrace$. Using this property, and the reciprocity relations (Eq. \ref{eqn:erecip}, Eq. \ref{eqn:mrecip}, and Eq. \ref{eqn:EMrecip}), we can derive the following equations:
 \begin{subequations}
 \label{eqn:trABrels}
 \begin{equation}
 \begin{split}
 \Re Tr & [\mathbf{\hat{n}}_2  (\mathbf{\tilde{r}})  \times \mu_2 \overline{\overline{G}}_e(\mathbf{\tilde{r}},\mathbf{r})]  \cdot 
     [\mathbf{\hat{n}}_1(\mathbf{r}) \times \varepsilon_1^* \overline{\overline{G}}_m^*(\mathbf{r},\mathbf{\tilde{r}})] = \\
   &  \Re Tr   [\mathbf{\hat{n}}_1(\mathbf{r}) \times \mu_1 \overline{\overline{G}}_e(\mathbf{r},\mathbf{\tilde{r}})] \cdot [\mathbf{\hat{n}}_2(\mathbf{\tilde{r}}) \times \varepsilon^*_2 \overline{\overline{G}}_m^*(\mathbf{\tilde{r}},\mathbf{r})]
 \end{split}
 \end{equation}
 \begin{equation}
  \label{eqn:trABrelsEM}
  \begin{split}
 \Re Tr & [\mathbf{\hat{n}}_2(\mathbf{\tilde{r}})  \times  \overline{\overline{G}}_E(\mathbf{\tilde{r}},\mathbf{r})] \cdot 
  [\mathbf{\hat{n}}_1(\mathbf{r}) \times \overline{\overline{G}}^*_E(\mathbf{r},\mathbf{\tilde{r}})]  =\\
& \Re Tr   [\mathbf{\hat{n}}_1(\mathbf{r}) \times \overline{\overline{G}}_M(\mathbf{r},\mathbf{\tilde{r}})] \cdot [\mathbf{\hat{n}}_2(\mathbf{\tilde{r}}) \times  \overline{\overline{G}}_M^*(\mathbf{\tilde{r}},\mathbf{r})]
 \end{split}
 \end{equation}
 \end{subequations}
 With the aid of Eq. \ref{eqn:trABrels}, $T^{e}_{1 \rightarrow 2}(\omega)$  can also be shown to be:
\begin{equation}
\label{eqn:generalizedtransmissivity2}
\begin{split}
 T^{e}_{1 \rightarrow 2}(\omega) & =  2\Re Tr \oint\limits_{S_1}d\mathbf{r} \oint\limits_{S_2}  d\mathbf{\tilde{r}} \bigg[ \frac{\omega^2}{c^2} \times \\
& [\mathbf{\hat{n}}_1(\mathbf{r}) \times \mu_1 \overline{\overline{G}}_e(\mathbf{r},\mathbf{\tilde{r}})] \cdot [\mathbf{\hat{n}}_2(\mathbf{\tilde{r}}) \times \varepsilon^*_2 \overline{\overline{G}}_m^*(\mathbf{\tilde{r}},\mathbf{r})]  \\
&+  [\mathbf{\hat{n}}_1(\mathbf{r}) \times \overline{\overline{G}}_M(\mathbf{r},\mathbf{\tilde{r}})] \cdot [\mathbf{\hat{n}}_2(\mathbf{\tilde{r}}) \times  \overline{\overline{G}}_M^*(\mathbf{\tilde{r}},\mathbf{r})]   \bigg] 
\end{split}
\end{equation}
The generalized transmissivity from object 2 to object 1, $ T^{e}_{2 \rightarrow 1}(\omega) $ can be determined from Eq. \ref{eqn:generalizedtransmissivity2} (or Eq. \ref{eqn:generalizedtransmissivity1}) by interchanging the subscripts 1 and 2 ($ \mathbf{r} \in S_1$ and $ \mathbf{\tilde{r}} \in S_2 $ are dummy variables and do not affect the value of the double integral).  $ T^{e}_{2 \rightarrow 1}(\omega) $ is given by:
\begin{equation}
\label{eqn:generalizedtransmissivity2to1}
\begin{split}
 T^{e}_{2 \rightarrow 1}(\omega)  & =  2\Re Tr \oint\limits_{S_2}  d\mathbf{\tilde{r}} \oint\limits_{S_1}d\mathbf{r}   \bigg[ \frac{\omega^2}{c^2} \times \\
&  [\mathbf{\hat{n}}_2(\mathbf{\tilde{r}}) \times \mu_2 \overline{\overline{G}}_e(\mathbf{\tilde{r}},\mathbf{r})] \cdot [\mathbf{\hat{n}}_1(\mathbf{r}) \times \varepsilon^*_1 \overline{\overline{G}}_m^*(\mathbf{r},\mathbf{\tilde{r}})]  \\
& + [\mathbf{\hat{n}}_2(\mathbf{\tilde{r}}) \times \overline{\overline{G}}_M(\mathbf{\tilde{r}},\mathbf{r})] \cdot [\mathbf{\hat{n}}_1(\mathbf{r}) \times  \overline{\overline{G}}_M^*(\mathbf{r},\mathbf{\tilde{r}})]   \bigg] 
\end{split}
\end{equation}
That the expressions for generalized transmissivity derived earlier (Eq. \ref{eqn:generalizedtransmissivity1} or Eq. \ref{eqn:generalizedtransmissivity2}) satisfy the principle of reciprocity in thermal radiative transfer, i.e. $ T^{e}_{2 \rightarrow 1}(\omega) = T^{e}_{1 \rightarrow 2}(\omega) $, can be established by using Eq.  \ref{eqn:trABrelsEM} to modify the expression for $ T ^{e}_{2 \rightarrow 1}(\omega) $ as follows:
\begin{equation}
\label{eqn:T12T21equality}
\begin{split}
 T^{e}_{2 \rightarrow 1}(\omega)  & =  2\Re Tr \oint\limits_{S_2}  d\mathbf{\tilde{r}} \oint\limits_{S_1}d\mathbf{r}  \bigg[   \frac{\omega^2}{c^2}  \times \\
& [\mathbf{\hat{n}}_2(\mathbf{\tilde{r}}) \times \mu_2 \overline{\overline{G}}_e(\mathbf{\tilde{r}},\mathbf{r})] \cdot [\mathbf{\hat{n}}_1(\mathbf{r}) \times \varepsilon^*_1 \overline{\overline{G}}_m^*(\mathbf{r},\mathbf{\tilde{r}})]  \\
& + [\mathbf{\hat{n}}_1(\mathbf{r})  \times \overline{\overline{G}}_E(\mathbf{r},\mathbf{\tilde{r}})] \cdot [\mathbf{\hat{n}}_2 (\mathbf{\tilde{r}}) \times  \overline{\overline{G}}_E^*(\mathbf{\tilde{r}},\mathbf{r})]   \bigg] \\
 & = T^{e}_{1 \rightarrow 2}(\omega) \quad  \text{ (Eq. \ref{eqn:generalizedtransmissivity1})}
\end{split}
\end{equation}

Though expressions for $ T^{e}_{1 \rightarrow 2}(\omega) $ in Eq. \ref{eqn:generalizedtransmissivity1} and Eq. \ref{eqn:generalizedtransmissivity2} are surface integrals, they are in fact derived from a volumetric integral over $ V_1 $ (Eq. \ref{eqn:eqhp*}). Similarly, energy emission from the object $ V_l $ ($ l=1,2,\cdots,N $) is derived from a volumetric integration over $ V_l $. For this reason, the formulae derived for $ T^{e}_{1 \rightarrow 2}(\omega) $ and $ T^{e}_{2 \rightarrow 1}(\omega) $ (Eq. \ref{eqn:generalizedtransmissivity1}, Eq. \ref{eqn:generalizedtransmissivity2}, or Eq. \ref{eqn:T12T21equality}) can be described as ``interior formulae.'' The integration over $ V_1 $ (for $ T^{e}_{1 \rightarrow 2}(\omega) $) or $ V_2 $ (for $ T^{e}_{2 \rightarrow 1}(\omega) $)  is made explicit by the presence of $ \varepsilon_1,\mu_2 $ (in Eq. \ref{eqn:generalizedtransmissivity1}) or $ \varepsilon_2,\mu_1 $ (Eq. \ref{eqn:generalizedtransmissivity2}, Eq. \ref{eqn:T12T21equality}). The corresponding ``exterior formula'' should not involve, or appear not to involve, any of these properties in the formula for transmissivity. 
The exterior formula for $ T^{e}_{1 \rightarrow 2} $ can be derived by using the boundary conditions (Eq. \ref{eqn:ebc}-Eq. \ref{eqn:Mbc}) and converting Eq. \ref{eqn:generalizedtransmissivity2} into the following equation:
\begin{equation}
\label{eqn:generalizedtransmissivityExterior}
\begin{split}
 T^{e}_{1 \rightarrow 2}(\omega)  & = 2\Re Tr \oint\limits_{S_1}d\mathbf{r} \oint\limits_{S_2}  d\mathbf{\tilde{r}} \bigg[ \frac{\omega^2}{c^2} \times \\
& [\mathbf{\hat{n}}_1(\mathbf{r}) \times \mu_h \overline{\overline{G}}_e(\mathbf{r},\mathbf{\tilde{r}})] \cdot [\mathbf{\hat{n}}_2(\mathbf{\tilde{r}}) \times \varepsilon^*_h \overline{\overline{G}}_m^*(\mathbf{\tilde{r}},\mathbf{r})]  \\
&+  [\mathbf{\hat{n}}_1(\mathbf{r}) \times \overline{\overline{G}}_M(\mathbf{r},\mathbf{\tilde{r}})] \cdot [\mathbf{\hat{n}}_2(\mathbf{\tilde{r}}) \times  \overline{\overline{G}}_M^*(\mathbf{\tilde{r}},\mathbf{r})]   \bigg] 
\end{split}
\end{equation}
There is a correspondence between the ``direct'' and ``indirect'' methods \cite{Narayanaswamy04A,narayanaswamy05a,wang2011direct} and the ``interior formula'' and ``exterior formula''  derived above. It can be shown that the ``exterior formula'' is a generalization of the ``indirect'' method to include problems of near-field thermal radiative energy transfer between two objects, in addition to the calculation of thermal emission from objects for which it is currently used. This correspondence will be undertaken in a future work and is not pursued any further in this paper.


Since $T^{e}_{2 \rightarrow 1}(\omega) = T^{e}_{1\rightarrow 2} (\omega)$, the net radiative exchange between objects 1 and 2, $ Q_{1,2} $, is given by:
\begin{equation}
\label{eqn:Q12gentrans}
Q_{1,2} = \int\limits_{0}^{\infty} \frac{d\omega}{2\pi} \left[ \Theta(\omega,T_1)-\Theta(\omega,T_2)\right]  T^{e}_{1\rightarrow 2}(\omega)
\end{equation}
From Eq. \ref{eqn:Q12gentrans}, a linearized conductance for radiative transfer between objects 1 and 2 can be defined as: 
\begin{equation}
\label{eqn:genconductance}
G^e_{1,2}(T)=\lim\limits_{T_1,T_2 \rightarrow T} \frac{Q_{1,2}}{T_1 - T_2} = \int\limits_{0}^{\infty} \frac{d\omega}{2\pi} \frac{\partial \Theta}{\partial T}  T^{e}_{1\rightarrow 2}(\omega)
\end{equation}

\subsection{Generalized transmissivity for radiative momentum transfer?}
Though expressions for generalized transmissivity  in terms of DGFs (Eq. \ref{eqn:generalizedtransmissivity1}, Eq. \ref{eqn:generalizedtransmissivity2}, Eq. \ref{eqn:T12T21equality}, and Eq. \ref{eqn:generalizedtransmissivityExterior}) have been derived for energy transfer, we have been unable to obtain equivalent expressions for generalized (vectorial) transmissivity or conductance of thermal non-equilibrium momentum transfer. Why this is so can be explained by considering the nature of the Poynting vector and the electromagnetic stress tensor. It is a well-known property of Maxwell's equations that the electric field at any location due to sources within a particular object, for instance object 1 in Fig. \ref{fig:generalinteraction}, can be expressed in terms of surface integrals of tangential electric and magnetic fields on the surface of that object. This property of electromagnetic fields forms the basis for the boundary element method for numerical solution of electromagnetic scattering problems. The normal component of the Poynting vector on the surface of object 2, $  \left( \mathbf{E}(\mathbf{\tilde{r}},t) \times \mathbf{H}(\mathbf{\tilde{r}},t)  \right) \cdot \mathbf{n}_2(\mathbf{\tilde{r}})$, has an additional property that it can be written as  $ \left[ \left(\mathbf{n}_2(\mathbf{\tilde{r}}) \times \mathbf{E}(\mathbf{\tilde{r}},t) \right) \times \left(\mathbf{n}_2(\mathbf{\tilde{r}}) \times \mathbf{H}(\mathbf{\tilde{r}},t)  \right)\right] \cdot \mathbf{n}_2(\mathbf{\tilde{r}})$. This ensures that the \textit{radiative heat transfer between objects 1 and 2 can be expressed in terms of tangential electric and magnetic fields on the surfaces of both objects}. This property of radiative heat transfer is reflected in the different expressions for $ T^{e}_{1 \rightarrow 2}(\omega) $  because they contain only tangential components of the DGFs on the surfaces of both objects. However, the electric and magnetic stress tensors in Eq. \ref{eqn:efieldstresstensor} and Eq. \ref{eqn:mfieldstresstensor} do not share this property. The force exerted by object 1 on object 2 requires knowledge of not only the tangential components of $ \mathbf{E} $ and $ \mathbf{H} $ on the surface of object 2 but also the normal components. While our inability to deduce an appropriate form for the transmissivity for momentum transfer does not mean that such a transmissivity does not exist, Eq. \ref{eqn:eecorrsurf}, and Eq. \ref{eqn:hhcorrsurf} can still be used to determine $ \overline{\overline{\sigma}}^e(\mathbf{r}) $ (Eq. \ref{eqn:efieldstresstensor}) and $ \overline{\overline{\sigma}}^m(\mathbf{r}) $ (Eq. \ref{eqn:mfieldstresstensor}) for specific geometric configuration of objects, from which thermal non-equilibrium van der Waals forces between objects can be computed.

\section{\label{sec:specgeom}Application to specific geometries or properties}
\subsection{\label{sec:parallelhalfspaces}Planar multilayered media}
\begin{figure}
\center
\includegraphics[width=3.3in]{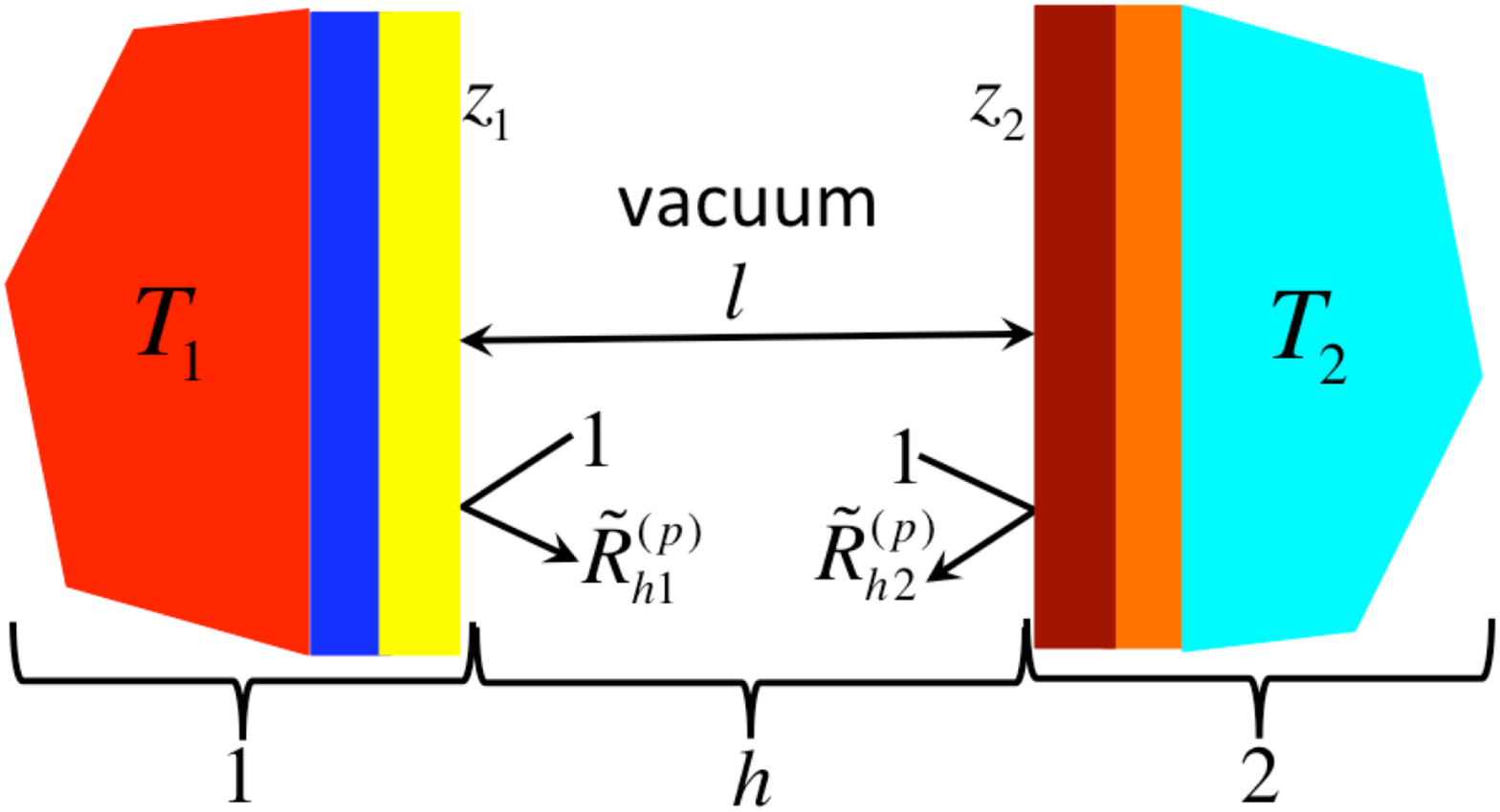}
\caption{\label{fig:twohalfspaces}Two multilayered half spaces at $ T_1 $ and $ T_2 $ separated by a vacuum gap.}
\end{figure}

Since theoretical analysis of radiative transfer and thermal non-equilibrium van der Waals forces between planar multilayered objects have been published in literature, the expressions derived in Sec. \ref{sec:fesif} are applied to objects shown in Fig. \ref{fig:twohalfspaces} and the resultant expressions compared with those in literature. The half spaces in Fig.  \ref{fig:twohalfspaces}, marked 1 and 2, can be homogeneous materials or can be composed of planar multilayer films. The only requirement is that the temperature gradients within 1 and 2 are negligible enough that they can be approximated as thermal reservoirs at temperatures $ T_1 $ and $ T_2 $. The vacuum layer of thickness $ l $ separating the two objects is equivalent to the host medium in Fig. \ref{fig:generalinteraction}. The planar media lie in the $ x-y $ plane and the unit vector in the $ z $ direction, $ \hat{z} $, is directed from object 1 to object 2. The interfaces of objects 1 and 2 with vacuum are at coordinates $ z=z_1 $ and $ z=z_2 $ ($ |z_2-z_1|=l $). The polarization dependent reflection coefficient of electromagnetic plane wave originating in vacuum and incident at the surface of half space 1 (in the absence of half space 2) is denoted as $ \widetilde{R}_{h1}^{(\mu)} $, where $\mu=s \text{ (transverse electric)}, \text{ } p \text{ (transverse magnetic)} $. A similar reflection coefficient for waves incident on half space 2, in the absence of half space 1, is denoted $ \widetilde{R}_{h2}^{(\mu)} $. $ \overline{\overline{G}}_o(\mathbf{r},\mathbf{\tilde{r}}) $ for any two locations $ \mathbf{r} $ and $ \mathbf{\tilde{r}} $ within the vacuum layer is given by:
\begin{equation}
\label{eqn:GoPlanar}
\begin{split}
  \overline{\overline{G}}_o  (\mathbf{r}, & \mathbf{\tilde{r}})  =  \frac{i}{4\pi}\int  \frac{dk_{\rho} k_{\rho}}{k_{hz}}  e^{i\mathbf{k}_{\rho} \cdot (\bm{\rho}-\bm{\rho}')} \times \\
& \sum\limits_{\mu=s,p} 
\begin{cases}
\hat{x}^{(\mu)}(+ k_{hz}) \hat{x}^{(\mu)}(+k_{hz})e^{i( z- z')k_{hz}} &\mbox{if } z>z' \\
\hat{x}^{(\mu)}(- k_{hz}) \hat{x}^{(\mu)}(-k_{hz})e^{i( z'- z)k_{hz}} &\mbox{if } z<z'
\end{cases}
\end{split}
\end{equation}
where $ \mathbf{k}_{\bm{\rho}} = k_{\rho} \hat{k}_{\rho}= k_x \hat{x} + k_y \hat{y}$, $ \hat{x}^{(s)}(\pm k_{hz}) = \hat{k}_{\rho} \times \hat{z} = (k_y \hat{x} - k_x \hat{y})/k_{\rho} $, $ \hat{x}^{(p)}(\pm k_{hz}) = (\mp k_{hz} \hat{k}_{ \bm{\rho}}+k_{\rho} \hat{z})/k_h $, and $ k_{hz}^2+k_{\rho}^2 = k_h^2 $. For vacuum, $ \mu_h = \varepsilon_h = 1 $ and $ k_h = \omega/c $. The scattered DGF, 
$ \overline{\overline{G}}_e^{(sc)}(\mathbf{r},\mathbf{\tilde{r}}) $,  is given by:
\begin{equation}
\label{eqn:GscPlanar}
\begin{split}
\overline{\overline{G}}_e^{(sc)}  & (\mathbf{r},\mathbf{\tilde{r}})  =  \frac{i}{4\pi}\int \frac{dk_{\rho} k_{\rho}}{k_{hz}}  \sum\limits_{\mu=s,p} \frac{e^{i\mathbf{k}_{\rho} \cdot (\bm{\rho}-\bm{\rho}')}}{D^{(\mu)}} \times \\
& \sum\limits_{\nu=\pm 1}\sum\limits_{\xi=\pm 1} C^{(\mu)}_{\nu,\xi} \hat{x}^{(\mu)}(\nu k_{hz}) \hat{x}^{(\mu)}(\xi k_{hz})e^{i(\nu z-\xi z')k_{hz}},
\end{split}
\end{equation}
where 
\begin{equation}
D^{(\mu)} = 1 - \widetilde{R}_{h1}^{(\mu)} \widetilde{R}_{h2}^{(\mu)} e^{i2k_{hz}l} 
\end{equation}
and
\begin{equation}
C^{(\mu)}_{\nu,\xi} = 
\begin{cases}
\displaystyle \widetilde{R}_{h1}^{(\mu)} \widetilde{R}_{h2}^{(\mu)} e^{i2k_{hz}l} &\mbox{if } \nu=\xi \\
\displaystyle \widetilde{R}_{h2}^{(\mu)} e^{i2k_{hz}z_2} &\mbox{if } \nu=-1,\xi=1\\
\displaystyle \widetilde{R}_{h1}^{(\mu)} e^{-i2k_{hz}z_1} &\mbox{if } \nu=1,\xi=-1
\end{cases}
\end{equation}
The total DGF, $ \overline{\overline{G}}_e  (\mathbf{r},\mathbf{\tilde{r}}) $, is given by $ \overline{\overline{G}}_o  (\mathbf{r},\mathbf{\tilde{r}}) + \overline{\overline{G}}_e^{(sc)}  (\mathbf{r},\mathbf{\tilde{r}}) $. 
The reflection coefficients $ \widetilde{R}_{h1}^{(\mu)} $ and $ \widetilde{R}_{h2}^{(\mu)} $ at the interface between two homogeneous media are given by the usual Fresnel reflection coefficients. For multilayered media, they can be computed using the transfer matrix method or by using recursion relations \cite{chew95a}. Since both $ \mathbf{r} $ and $ \mathbf{\tilde{r}} $ are within the same layer, $  \overline{\overline{G}}_m  (\mathbf{r},\mathbf{\tilde{r}})  $ and $  \overline{\overline{G}}_M  (\mathbf{r},\mathbf{\tilde{r}})  $ can be determined by simply replacing all occurrences of $ \varepsilon(\omega) $ by the corresponding $ \mu(\omega) $ and vice versa. Explicit expressions for $ \overline{\overline{G}}_m^{(sc)}  (\mathbf{r},\mathbf{\tilde{r}}) $, $ \overline{\overline{G}}_E^{(sc)}  (\mathbf{r},\mathbf{\tilde{r}}) $, and $ \overline{\overline{G}}_M^{(sc)}  (\mathbf{r},\mathbf{\tilde{r}}) $ are given below:
\begin{equation}
\begin{split}
  \overline{\overline{G}}_m^{(sc)}  (\mathbf{r}, & \mathbf{\tilde{r}}) =  \frac{i}{4\pi}\int  \frac{dk_{\rho} k_{\rho}}{k_{hz}}  \sum\limits_{\mu=s,p} \frac{e^{i\mathbf{k}_{\rho} \cdot (\bm{\rho}-\bm{\rho}')}}{D^{(\mu')}} \times \\
& \sum\limits_{\nu=\pm 1}\sum\limits_{\xi=\pm 1} C^{(\mu')}_{\nu,\xi} \hat{x}^{(\mu)}(\nu k_{hz}) \hat{x}^{(\mu)}(\xi k_{hz})e^{i(\nu z-\xi z')k_{hz}},
\end{split}
\end{equation}
where $ \mu'=p $ if $ \mu = s $ and $ \mu'=s $ if $ \mu = p $. Defining $\hat k_h^{(\pm)}=(k_{\rho}\hat k_{\rho} \pm k_{hz}\hat z)/k_h$, we have the following relations: $\hat k_{h}^{(\pm)} \times \hat{x}^{(s)}(\pm k_{hz})=- \hat{x}^{(p)}(\pm k_{hz})$ and $\hat k_{h}^{(\pm)} \times \hat{x}^{(p)}(\pm k_{hz})=\hat{x}^{(s)}(\pm k_{hz})$. Using these relations, we obtain: 
\begin{equation}
\begin{split}
  \overline{\overline{G}}_E^{(sc)}  &  (\mathbf{r},\mathbf{\tilde{r}}) =  \frac{-1}{4\pi}\int  \frac{dk_{\rho} k_{\rho}}{k_{hz}}  k_{h} \sum\limits_{\mu=s,p} \frac{e^{i\mathbf{k}_{\rho} \cdot (\bm{\rho}-\bm{\rho}')}}{D^{(\mu)}} \times \\
& \sum\limits_{\nu=\pm 1}\sum\limits_{\xi=\pm 1} (-1)^{\beta} C^{(\mu)}_{\nu,\xi} \hat{x}^{(\mu')}(\nu k_{hz}) \hat{x}^{(\mu)}(\xi k_{hz})e^{i(\nu z-\xi z')k_{hz}},
\end{split}
\end{equation}
\begin{equation}
\begin{split}
  \overline{\overline{G}}_M^{(sc)} & (\mathbf{r},\mathbf{\tilde{r}}) =  \frac{-1}{4\pi}\int  \frac{dk_{\rho} k_{\rho}}{k_{hz}}  k_{h} \sum\limits_{\mu=s,p} \frac{e^{i\mathbf{k}_{\rho} \cdot (\bm{\rho}-\bm{\rho}')}}{D^{(\mu')}} \times \\
& \sum\limits_{\nu=\pm 1}\sum\limits_{\xi=\pm 1} (-1)^{\beta} C^{(\mu')}_{\nu,\xi} \hat{x}^{(\mu')}(\nu k_{hz}) \hat{x}^{(\mu)}(\xi k_{hz})e^{i(\nu z-\xi z')k_{hz}},
\end{split}
\end{equation}
where $ \beta=0 \mbox{ if } \mu=p $ and $ \beta=1 \mbox{ if } \mu=s $. 

A key distinction between the ``direct'' and ``exterior'' methods can be illustrated through the example being considered here. Let us assume that the half space $ L $ is composed of multiple planar films. In the direct method \cite{Narayanaswamy04A,narayanaswamy05a}, the contribution of each layer is evaluated separately and added subsequently to determine the heat flux in the vacuum layer due to half space $ L $. This requires finding the DGFs when $ \mathbf{\tilde{r}} $ belongs to the vacuum layer and $ \mathbf{r} $ lies in each of the thin films that makes up the half spaces $ L $. On the other hand, the exterior method, which is used to derive Eq. \ref{eqn:twohsenergy} and Eq. \ref{eqn:twohsforce}, requires only  knowledge of the DGFs when $ \mathbf{r} $ and $ \mathbf{\tilde{r}} $ belong to the vacuum layer. Though one can derive the same results, but with more algebraic manipulations, for layered media, the ease of using the exterior method should become  apparent when one tries to model near-field radiative transfer in more complicated geometries, for instance between two coated spheres.

\subsubsection{\label{sec:twohsenergy}Two parallel half spaces: Radiative transfer}
 For the two half spaces in Fig. \ref{fig:twohalfspaces}, the surface normal vectors $ \mathbf{\hat{n}}(\mathbf{r}) $ and $ \mathbf{\hat{n}}(\mathbf{\tilde{r}}) $ in Eq. \ref{eqn:generalizedtransmissivityExterior} are given by $ \hat{z} $ and $ -\hat{z} $ respectively. $ \overline{\overline{G}}_e  (\mathbf{r},\mathbf{\tilde{r}}) $, $ \overline{\overline{G}}_m  (\mathbf{r},\mathbf{\tilde{r}}) $, $ \overline{\overline{G}}_E  (\mathbf{r},\mathbf{\tilde{r}}) $, and $ \overline{\overline{G}}_M  (\mathbf{r},\mathbf{\tilde{r}}) $ derived using Eq. \ref{eqn:GoPlanar} and Eq. \ref{eqn:GscPlanar} are substituted into Eq. \ref{eqn:generalizedtransmissivityExterior} to obtain, $ T^{e,pp}_{1 \rightarrow 2} $, the generalized transmissivity of energy transfer between planar half spaces (the superscript $ pp $ is short for ``planar-planar'' and is used to indicate the type of objects). After some manipulations, $ T^{e,pp}_{1 \rightarrow 2} $ can be shown to be:
\begin{equation}
\label{eqn:twohsenergy}
\begin{split}
T^{e,pp}_{1 \rightarrow 2}(\omega) =&  \int\limits_0^{\omega/c}  \frac{k_{\rho} dk_{\rho} }{2\pi}    \sum\limits_{\mu=s,p} \frac{(1-\lvert \widetilde{R}_{h1}^{(\mu)} \rvert^2)(1-\lvert \widetilde{R}_{h2}^{(\mu)} \rvert^2)}{\lvert 1 - \widetilde{R}_{h1}^{(\mu)} \widetilde{R}_{h2}^{(\mu)} e^{i2k_{hz} l} \rvert^2} + \\
& \int\limits^{\infty}_{\omega/c}  \frac{k_{\rho} dk_{\rho} }{2\pi}    \sum\limits_{\mu=s,p} \frac{4\Im (\widetilde{R}_{h1}^{(\mu)}) \Im (\widetilde{R}_{h2}^{(\mu)}) e^{-2 |k_{hz}| l}}{\lvert 1 - \widetilde{R}_{h1}^{(\mu)} \widetilde{R}_{h2}^{(\mu)} e^{-2|k_{hz}| l} \rvert^2}
\end{split}
\end{equation}
This expression for $ T^{e,pp}_{1\rightarrow 2}(\omega) $ is in agreement with expressions for transmissivity of energy transfer across half spaces \cite{biehs2010mesoscopic}. For the case of two homogeneous half spaces, we have also confirmed that volume integral expression (Eq. \ref{eqn:eqhp*}) as well as surface integral expressions (Eq. \ref{eqn:generalizedtransmissivity1}, Eq. \ref{eqn:generalizedtransmissivityExterior}) yield the same result. Two interesting features of Eq. \ref{eqn:twohsenergy} need to be emphasized: (1) Eq. \ref{eqn:twohsenergy} is valid for energy transfer not just between  two homogeneous half spaces but also between two half spaces comprising planar thin films, (2) It is valid for isotropic materials with electric as well as magnetic polarizabilities, i.e., with frequency dependent $ \varepsilon $ and $ \mu $. 

\subsubsection{\label{sec:twohsforce}Two parallel half spaces: Non-equilibrium pressure}
Since we have not been able to derive a generalized transmissivity for momentum transfer, we use Eq. \ref{eqn:efieldstresstensor} and Eq. \ref{eqn:mfieldstresstensor} to derive the van der Waals pressure in the vacuum gap. Because the film is perpendicular to the $ z $ direction, the van der Waals pressure is given by the $ zz $ component of the stress tensor, $ \sigma_{zz} $. Using the expressions for  $ \overline{\overline{G}}_e  (\mathbf{r},\mathbf{\tilde{r}}) $ and $ \overline{\overline{G}}_E  (\mathbf{r},\mathbf{\tilde{r}}) $ in Eq. \ref{eqn:eecorrsurf} and $ \overline{\overline{G}}_m  (\mathbf{r},\mathbf{\tilde{r}}) $ and $ \overline{\overline{G}}_M  (\mathbf{r},\mathbf{\tilde{r}}) $ in Eq.  \ref{eqn:hhcorrsurf}, a \textit{generalized transmissivity for momentum flux} from $ L $ to $ R $ is given by:
\begin{equation}
\label{eqn:twohsforce}
\begin{split}
T^{m,pp}_{1 \rightarrow 2}(\omega) =&  -\int\limits_0^{\omega/c}  \frac{k_{\rho} dk_{\rho} }{2\pi}  \frac{ k_{hz}}{\omega}   \sum\limits_{\mu=s,p} \frac{(1-\lvert \widetilde{R}_{h1}^{(\mu)} \rvert^2)(1+\lvert \widetilde{R}_{h2}^{(\mu)} \rvert^2)}{\lvert 1 - \widetilde{R}_{h1}^{(\mu)} \widetilde{R}_{h2}^{(\mu)} e^{i2k_{hz} l} \rvert^2} + \\
& \int\limits^{\infty}_{\omega/c}  \frac{k_{\rho} dk_{\rho} }{2\pi}  \frac{ |k_{hz}|}{\omega}  \sum\limits_{\mu=s,p} \frac{4\Im (\widetilde{R}_{h1}^{(\mu)}) \Re (\widetilde{R}_{h2}^{(\mu)}) e^{-2 |k_{hz}| l}}{\lvert 1 - \widetilde{R}_{h1}^{(\mu)} \widetilde{R}_{h2}^{(\mu)} e^{-2|k_{hz}| l} \rvert^2}
\end{split}
\end{equation}
The superscript $ m $ in $T^{m,pp}_{1 \rightarrow 2}(\omega)$ stands for ``momentum.'' 
Using the same notation as Antezza et. al. \cite{antezza2008casimir}, the non-equilibrium pressure in the vacuum layer due to temperature $ T_L $ of half space $  L$ while $ T_R = 0 $ K, denoted by $ P_{neq}(T_L,T_R=0,l) $, is calculated using the formula $ P_{neq}(T_L,T_R=0,l) = \int\limits_0^{\infty} \frac{d\omega}{2\pi} \Theta(\omega,T_L) T^{m,pp}_{1 \rightarrow 2}(\omega)$. It is interesting to note that Eq. \ref{eqn:twohsforce}, which is valid for half spaces with arbitrary $ \varepsilon $ and $ \mu $, coincide with the expressions for non-equilibrium van der Waals pressure and radiative transfer derived in Ref. \cite{antezza2008casimir} and Ref. \cite{biehs2010mesoscopic} even though the authors of Ref. \cite{antezza2008casimir} and Ref. \cite{biehs2010mesoscopic} derived it only for the case when $ \mu = 1 $ everywhere. 

\subsection{\label{sec:bbagreement}Agreement with theory of blackbody radiative transfer}
Computing radiative transfer between two arbitrarily shaped isotropic objects using Eq. \ref{eqn:generalizedtransmissivity1}, Eq. \ref{eqn:generalizedtransmissivity2}, or Eq. \ref{eqn:generalizedtransmissivityExterior} is computationally involved because of the need to compute the appropriate DGFs. However, for one class of objects, namely blackbodies (or objects those can be approximated as blackbodies), the expression for generalized transmissivity derived here can be used to obtain useful results irrespective of the shape. Blackbody radiative transfer is derived from Planck's theory of blackbody radiation and Kirchoff's laws, both of which are consequences of thermodynamics applied to relatively simple electrodynamical systems (for example, photon gas in a piston with perfectly reflective walls). Using thermodynamic arguments, this idea is generalized to arbitrarily shaped objects to yield $Q_{1,2}^{bb}=A_1 F_{1,2} \sigma_{SB} \left(T_1^4 - T_2^4 \right)$, where $ \sigma_{SB} $ is the Stefan-Boltzmann constant, the superscript $ bb $ stands for ``blackbody'', and $ F_{1,2} $ is the view factor between objects 1 and 2. The view factor $F_{1,2}$ between the two objects in Fig. \ref{fig:generalinteraction} is given by: 
\begin{equation}
\label{eqn:viewfactorbb}
F_{1,2} = \frac{1}{A_1}\oint\limits_{S_1} d\mathbf{r} \oint\limits_{S_2} d\mathbf{\tilde{r}} \frac{\left(- \mathbf{n}_1 \cdot \mathbf{\hat{R}} \right)  \left( \mathbf{n}_2 \cdot \mathbf{\hat{R}}\right)}{\pi R^2},
\end{equation}

\noindent where $\mathbf{\hat{R}} = (\mathbf{r} - \mathbf{\tilde{r}})/| \mathbf{r} - \mathbf{\tilde{r}}|$, $\mathbf{R} =R\mathbf{\hat{R}} = \mathbf{r} - \mathbf{\tilde{r}}$,  $\mathbf{r} \in S_1$, and $\mathbf{\tilde{r}} \in S_2$, and $A_1$ is the area of $S_1$. 

A blackbody is one that absorbs all radiation incident on it and scatters none. For an object in vacuum ($ \varepsilon_h=\mu_h=1 $), this can be achieved by a region of space (the blackbody) with permittivity and  permeability given by $\varepsilon = 1+i \delta$, $\mu = 1+i\gamma $ such that $\delta, \gamma \rightarrow 0$, ensuring that there is no scattering by the object. The nominal dimension $L$ should be such that $\delta k L \gg 1$ or $\gamma  k L \gg 1$, where $k =  \omega/c$ and $c$ is the speed of light in vacuum, ensuring that all the radiation entering the object is absorbed. Because the properties of the objects differ infinitesimally from that of the host medium, scattering can effectively be neglected and the DGFs, $ \overline{\overline{G}}_e(\mathbf{r}, \mathbf{\tilde{r}}) $ and $ \overline{\overline{G}}_m(\mathbf{r}, \mathbf{\tilde{r}}) $, are simply given by the DGF in free space, which is:
\begin{equation}
\label{eqn:homodyade}
\begin{split}
 \overline{\overline{G}}_o(\mathbf{r}, \mathbf{\tilde{r}}) = \frac{e^{ik_h R}}{4\pi R}\Bigg[ & \mathbf{\hat{R}}\mathbf{\hat{R}} (-i\frac{2}{k_hR}+\frac{2}{k_h^2R^2}) + \\
& (\overline{\overline{I}}- \mathbf{\hat{R}}\mathbf{\hat{R}})
 \left(1+\frac{i}{k_hR} - \frac{1}{k_h^2R^2}\right) 
\Bigg]
\end{split}
\end{equation}

\noindent When the spacing between objects is large compared to the wavelength, Eq. \ref{eqn:homodyade} reduces to: 
\begin{equation}
\label{eqn:homodyadeapprox}
\overline{\overline{G}}_e(\mathbf{r}, \mathbf{\tilde{r}}) = \overline{\overline{G}}_m(\mathbf{r}, \mathbf{\tilde{r}}) = \overline{\overline{G}}_o(\mathbf{r}, \mathbf{\tilde{r}}) = \frac{e^{ik_hR}}{4\pi R}  (\overline{\overline{I}}- \mathbf{\hat{R}}\mathbf{\hat{R}}) 
\end{equation}

\noindent Similarly, $ \overline{\overline{G}}_E(\mathbf{r}, \mathbf{\tilde{r}}) $ and $ \overline{\overline{G}}_M(\mathbf{r}, \mathbf{\tilde{r}}) $ can be written as: 
\begin{equation} 
\label{eqn:homodyadecurlapprox}
 \nabla \times \overline{\overline{G}}_o(\mathbf{r}, \mathbf{\tilde{r}}) =\overline{\overline{G}}_O(\mathbf{r}, \mathbf{\tilde{r}}) = ik_h \frac{e^{ik_hR}}{4\pi R}  (\mathbf{\hat{R}} \times \overline{\overline{I}})
\end{equation}

To derive the generalized transmissivity between two blackbodies, the following derivations are useful:
\begin{equation}
\label{eqn:bbtransmissivity1}
\begin{split}
\Re Tr  &\big[(\mathbf{\hat{n}}_1(\mathbf{r})  \times  \mu_h \overline{\overline{G}}_e(\mathbf{r},\mathbf{\tilde{r}})) \cdot (\mathbf{\hat{n}}_2(\mathbf{\tilde{r}}) \times  \varepsilon_h \overline{\overline{G}}_m(\mathbf{\tilde{r}},\mathbf{r}))^* \big]  \\
=&\Re Tr  \big[(\mathbf{\hat{n}}_1(\mathbf{r})  \times  \overline{\overline{G}}_o(\mathbf{r},\mathbf{\tilde{r}})) \cdot (\mathbf{\hat{n}}_2(\mathbf{\tilde{r}}) \times  \overline{\overline{G}}_o(\mathbf{\tilde{r}},\mathbf{r}))^* \big]  \\
=& \frac{\big[\mathbf{\hat{n}}_1(\mathbf{r})  \times  (\overline{\overline{I}}- \mathbf{\hat{R}}\mathbf{\hat{R}})]_{ij}\big[\mathbf{\hat{n}}_2(\mathbf{\tilde{r}})  \times  (\overline{\overline{I}}- \mathbf{\hat{R}}\mathbf{\hat{R}})]_{ji}}{(4\pi R)^2} \\
=& \frac{\big[\mathbf{\hat{n}}_1(\mathbf{r})  \times  (\mathbf{\hat{\theta}}\mathbf{\hat{\theta}} + \mathbf{\hat{\phi}}\mathbf{\hat{\phi}})]_{ij}\big[\mathbf{\hat{n}}_2(\mathbf{\tilde{r}})  \times  (\mathbf{\hat{\theta}}\mathbf{\hat{\theta}} + \mathbf{\hat{\phi}}\mathbf{\hat{\phi}})]_{ji}}{(4\pi R)^2} \\
= & \frac{-2(\mathbf{\hat{n}}_1(\mathbf{r}) \cdot \mathbf{\hat{R}}) (\mathbf{\hat{n}}_2(\mathbf{\tilde{r}}) \cdot \mathbf{\hat{R}})}{(4\pi R)^2}
\end{split}
\end{equation}
\begin{equation}
\label{eqn:bbtransmissivity2}
\begin{split}
\Re  Tr   \big[&(\mathbf{\hat{n}}_1(\mathbf{r})  \times  \overline{\overline{G}}_M(\mathbf{r},\mathbf{\tilde{r}})) \cdot (\mathbf{\hat{n}}_2(\mathbf{\tilde{r}}) \times  \overline{\overline{G}}_M(\mathbf{\tilde{r}},\mathbf{r}))^* \big]  \\
=\Re & Tr   \big[(\mathbf{\hat{n}}_1(\mathbf{r})  \times  \overline{\overline{G}}_O(\mathbf{r},\mathbf{\tilde{r}})) \cdot (\mathbf{\hat{n}}_2(\mathbf{\tilde{r}}) \times  \overline{\overline{G}}_O(\mathbf{\tilde{r}},\mathbf{r}))^* \big]  \\
 = -&k_h^2 \frac{\big[\mathbf{\hat{n}}_1(\mathbf{r})  \times  (\mathbf{\hat{R}} \times \overline{\overline{I}})]_{ij}\big[\mathbf{\hat{n}}_2(\mathbf{\tilde{r}})  \times  (\mathbf{\hat{R}} \times \overline{\overline{I}})]_{ji}}{(4\pi R)^2} \\
=-&k_h^2 \frac{\big[\mathbf{\hat{n}}_1(\mathbf{r})  \times  (\mathbf{\hat{\phi}}\mathbf{\hat{\theta}} - \mathbf{\hat{\theta}}\mathbf{\hat{\phi}})]_{ij}\big[\mathbf{\hat{n}}_2(\mathbf{\tilde{r}})  \times  (\mathbf{\hat{\phi}}\mathbf{\hat{\theta}} - \mathbf{\hat{\theta}}\mathbf{\hat{\phi}})]_{ji}}{(4\pi R)^2} \\
= k&_h^2  \frac{-2(\mathbf{\hat{n}}_1(\mathbf{r}) \cdot \mathbf{\hat{R}}) (\mathbf{\hat{n}}_2(\mathbf{\tilde{r}}) \cdot \mathbf{\hat{R}})}{(4\pi R)^2}
\end{split}
\end{equation}
Substituting Eq. \ref{eqn:bbtransmissivity1} and Eq. \ref{eqn:bbtransmissivity2} in Eq. \ref{eqn:generalizedtransmissivityExterior}, the generalized transmissivity between two blackbodies in the far-field is:
\begin{equation}
\begin{split}
T_{1\rightarrow2}^{bb} (\omega) = & \frac{\omega^2}{c^2}\oint\limits_{S_1} d\mathbf{r}  \oint\limits_{S_2} d\mathbf{\tilde{r}} \frac{\left(- \mathbf{n}_1(\mathbf{r}) \cdot \mathbf{\hat{R}} \right)  \left( \mathbf{n}_2(\mathbf{\tilde{r}}) \cdot \mathbf{\hat{R}}\right)}{2\pi^2 R^2}  \\
= & \frac{\omega^2}{2\pi c^2} A_1 F_{1,2}
\end{split}
\end{equation}
and $ Q_{1,2}^{bb} $ is given by:
\begin{equation}
\label{eqn:bbfinal}
\begin{split}
Q_{1,2}^{bb} = & \int\limits_{0}^{\infty} \frac{d\omega}{2\pi} \frac{\omega^2}{2\pi c^2} A_1 F_{1,2} [\Theta(\omega,T_1)-\Theta(\omega,T_2)]\\
= &  A_1 F_{1,2} \frac{\pi^2 k_B^4}{60 c^2 \hbar^3} (T_1^4 - T_2^4) \\
= &  A_1 F_{1,2} \sigma_{SB} (T_1^4 - T_2^4) 
\end{split}
\end{equation}
The agreement of obtained results in the examples of Sec. \ref{sec:parallelhalfspaces} with earlier published works \cite{biehs2010mesoscopic,antezza2008casimir} and the above derivation for the radiative heat transfer between blackbodies attest to the correctness  of our definitions of the generalized transmissivity (Eq. \ref{eqn:generalizedtransmissivity1}, Eq. \ref{eqn:generalizedtransmissivity2}, and Eq. \ref{eqn:generalizedtransmissivityExterior}). Extension of the proof given here to obtain Eq. \ref{eqn:bbfinal} to the case of gray body radiative transfer is not a simple one because we consider only specular reflection at surfaces. 

\section{\label{sec:summary}Summary}
In this paper, we have developed a dyadic Green's function formalism to determine radiative heat transfer  and non-equilibrium van der Waals/Casimir forces between objects of arbitrary shapes, sizes, and with frequency dependent dielectric permittivity and magnetic permeability. The cross-spectral densities of electromagnetic fields are necessary to evaluate Poynting vector and electromagnetic stress tensor from which radiative transfer and forces between objects can be evaluated. Using Rytov's fluctuational electrodynamics, expressions for cross-spectral densities in terms of volume integrals of products of DGFs are obtained. Green's identities for dyadic functions are then used to convert these volume integral expressions into surface integrals of products of tangential components of the DGFs on the surfaces of scatterers. The spectral radiative transfer between two objects is described in terms of a single quantity - the generalized transmissivity - which can be represented as a double integral of products of tangential components over the surfaces of the two objects. The spectral integral of $ T_{1 \rightarrow 2}^{e} (\omega)$ weighted by the temperature derivative of the Bose-Einstein function yields the thermal radiative conductance between the two objects. 
 In the geometric optics limit, the thermal radiative conductance between two blackbodies, as derived from the generalized transmissivity, is shown to agree with the predictions of the classical theory of radiative transfer. 

While many computational methods, such as finite element method, vector eigenfunction expansion, or T-matrix method, can be employed to compute $ T_{1 \rightarrow 2}^{e} (\omega)$, the surface integral expression in Eq. \ref{eqn:generalizedtransmissivityExterior} points towards the surface integral equation method (SIEM) as best-suited for the purpose. The main advantage of a surface integral equation based method is the potential reduction in computational cost due to restriction of the discretization domain to a surface rather than a volume. This advantage becomes more important as the size of the object becomes larger. After the submission of this work, we were made aware of a recently released free software titled SCUFF-EM (\textbf{S}urface \textbf{CU}rrents \text{F}ield \textbf{F}ormulation of \textbf{E}lectro-\textbf{M}agnetism) \footnote{It can be downloaded from \url{http://homerreid.ath.cx/scuff-EM/}} based on a surface integral formulation of the electromagnetic scattering problem \cite{reid2012fluctuating,rodriguez2012fluctuating}.    
The similarities between photon and phonon transport lead us to wonder whether such a surface integral formulation for phonon energy transport in mesoscale structures in the harmonic limit is also possible. 

This work has been funded partially by National Science Foundation through Grant CBET-0853723.

\bibliographystyle{model3-num-names}

\end{document}